\documentclass[universe,review,submit,pdftex,oneauthor]{Definitions/mdpi} 

\firstpage{1} 
\pubvolume{1}
\issuenum{1}
\articlenumber{0}
\pubyear{2026}
\copyrightyear{2026}
\datereceived{ } 
\daterevised{ } 
\dateaccepted{ } 
\datepublished{ } 

\let\linenumbers\relax
\usepackage{awesomebox}
\usepackage{natbib}
\graphicspath{{figures/}}

\Title{Observations of stellar feedback: tracers, diagnostics, measurements, and physical interpretation}

\Author{Anna F. McLeod $^{1,2}$\orcidA{}*}

\AuthorNames{Anna F. McLeod}

\address{%
$^{1}$Centre for Extragalactic Astronomy, Department of Physics, University of Durham, South Road, Durham DH1 3LE, UK\\
$^{2}$Institute for Computational Cosmology, Department of Physics, University of Durham, South Road, Durham DH1 3LE, UK
}

\corres{Correspondence: anna.mcleod@durham.ac.uk}

\abstract{Stellar feedback is central to galaxy evolution, yet the coupling of radiation, stellar winds, and supernovae to the multi-phase interstellar medium remains uncertain. This review assesses this as a problem of inference and frames the discussion around four qualitative axes of ignorance, that is: observational diagnostics are indirect; stellar population synthesis models need to be ground-truthed to be used to infer feedback quantities; recent high-redshift observations have opened a new regime that potentially requires re-calibrating; and stellar feedback and the affected environment are interdependent. The inference chain of feedback, which requires  conversions from observables to physical quantities and from those quantities to an interpretation of how energy and momentum are transferred to the surrounding medium, is examined across the principal observational diagnostics of stellar feedback: from ionised gas and mechanically-driven bubbles to giant molecular cloud lifetimes, galactic outflows, and Lyman-continuum escape, to address the overarching question of how much observations actually allow us to know about the link between the feedback-driving massive stars and the feedback-driven interstellar medium. Further, we discuss how the assumptions that go into stellar population synthesis models as well as environmental (e.g., metallicity, dust content) and galaxy properties (e.g., stellar mass, gas content) affect feedback measurements, and how recent high-redshift observations are testing calibrations established in the nearby Universe. The emerging picture is that the interpretation of a feedback measurement requires explicit control of the inference chain and of the physical context in which the measurement is made, and that feedback should consequently be {\it contextualised}, meaning that measurements must be interpreted in terms of the feedback mechanism, gas phase, environment, and spatial and temporal scale. Progress in the field will depend on measuring feedback-related quantities across the full range of star-forming systems, and, more importantly, on obtaining observations ("real" as well as synthetic from post-processed simulations) that match the same tracer and spatial scale across the same systems.}


\begin{document}


\section{Introduction}

\begin{figure}
    \centering
    \includegraphics[scale=0.5]{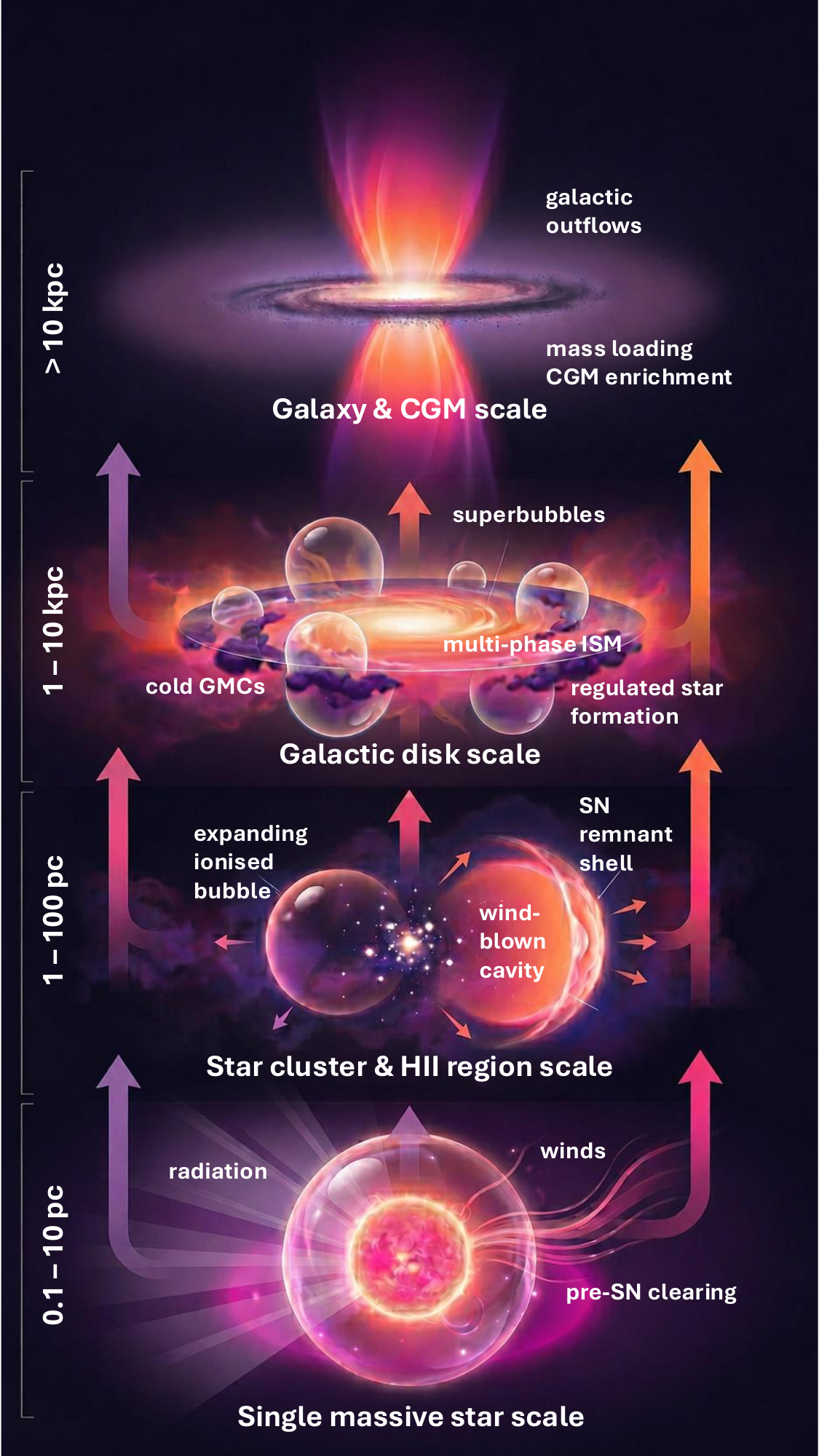}
    \caption{Schematic showing the scales and components of stellar feedback, from individual massive stars (bottom) to the kpc scales of galaxy disks, feedback-driven outflows, and the circumgalactic medium (top). Radiation and stellar winds start to inflate and clear HII regions before the first supernovae start contributing to dispersing the natal molecular material. The combined effect of stellar feedback affects the multi-phase ISM up to kpc scales by inflating superbubbles and driving galactic outflows, thus playing a role in the regulation of star formation and the re-distribution of chemically-enriched material.}
    \label{fig:ladder}
\end{figure}

Over the short (order-of-Myr) timescales relevant to star formation and the regulation of the interstellar medium (ISM), massive stars (M $>$ 8 M$_{\odot}$) dominate the radiative and mechanical energy input. By emitting photons that ionise and inject momentum into the surrounding ISM, launching powerful stellar winds, (a subset of) massive stars being progenitors to energetic core-collapse supernova (CCSN) events, and chemically enriching their surroundings, they affect the ISM from the sub-pc scales of individual stars to the kpc scales of the circumgalactic medium (CGM; see Figure~\ref{fig:ladder}). These processes, via which feedback couples to the ISM, are collectively referred to as stellar feedback, and play a central role in the baryon cycle of galaxies \citep{naab2017}. 

From small to large scales, stellar feedback is responsible for the photoevaporation of protoplanetary disks, the expansion of HII regions, the disruption of molecular clouds, the launching of galactic winds (outflows) which in low-mass galaxies are able to alter the underlying dark matter distribution by removing baryonic matter from the potential. The importance of stellar feedback for galaxy evolution is demonstrated by the inability of cosmological simulations lacking efficient feedback to reproduce the observed galaxy population. Such simulations convert too large a fraction of baryons into stars, yielding galaxies with stellar masses substantially larger than observed and star formation histories (SFHs) that are inconsistent with observations, with morphologies bearing little resemblance to the Milky Way or its Local Group neighbors \citep{schaye2015,scannapieco2012}. The inclusion of stellar feedback — resulting in the net suppression of star formation by injection of turbulence, dispersal of molecular clouds, and the launching of galactic-scale outflows — is therefore a necessity for any plausible model of galaxy formation and evolution \citep{schaye2025,balogh2001}. Yet, despite decades of progress, the efficiency with which feedback couples to the multi-phase ISM across the full range of galactic environments, and how this may have changed with cosmic time, remains uncertain.

This review focuses on four qualitative axes of ignorance that limit the predictive power of current observational feedback studies and that the community must address to drive progress that is not merely incremental.

\vspace{0.2cm}

\noindent {\bf Axis I: Observational diagnostics require calibration and constraints.} Quantifying the  dynamical coupling efficiency of stellar feedback to the multi-phase ISM is based on the inference of feedback-related quantities. The observables we use to derive these — emission line luminosities, line profiles, expansion velocities, mass outflow rates — are a convolution of stellar physics with ISM geometry, density structure, ionisation state, and kinematic complexity, and are furthermore averaged along the line of sight and over timescales set by both stellar evolution and the response of the affected gas. Emission line-based inferences inherit additional systematic uncertainty from dust attenuation, sampling of the initial mass function (IMF), and spatial resolution \citep{kennicutt2012}, to name a few. Converting observed photons into quantities such as energy injection rates, mass outflow rates, or molecular cloud dispersal timescales therefore requires a chain of assumptions, each of which can propagate into the final inferred quantity. Further, while feedback studies are now entering the statistical regime enabled by integral field surveys at cloud scale across entire galaxies, most remain limited by spatial resolution and limited probed environmental diversity \citep{emsellem2022,barnes2021,belfiore2022}. 

\vspace{0.2cm}

\noindent {\bf Axis II: Stellar population models carry unresolved uncertainties.} Deriving an ionising photon rate, a SFH, or the supernova rate of the underlying stellar population from an observed spectral energy distribution (SED) or emission line spectrum, requires population synthesis models. These come with internal degeneracies and depend on the details of stellar evolution (wind mass-loss prescriptions, binary evolution, and the adopted stellar atmosphere and interior tracks \citep{leitherer1999,eldridge2017,conroy2013}), as well as on the properties of the assumed IMF. Advancements in these areas, such as the downward revisions to empirical and predicted stellar wind mass-loss rates that can amount to factors of several \citep{puls2008,vink2011}, or the recognition that more than 70\% of massive stars undergo binary mass exchange over their lifetimes \citep{sana2012}, change inferred ionising flux, supernova rates, and mechanical yields in population synthesis \citep{eldridge2017,agrawal2022}. These revisions have yet to be propagated into galaxy-scale models and simulations, enabling the validation of empirically-derived feedback quantities.

\vspace{0.2cm}

\noindent {\bf Axis III: High-z opens qualitatively new physical regimes.} High-redshift galaxies are characterised by lower metallicities, higher gas fractions, higher ISM pressures, and more extreme specific star formation rates (sSFRs) than the low-redshift samples on which most diagnostic calibrations were developed \citep{maiolino2019,tacconi2020}. Photometry from the James Webb Space Telescope ({\it JWST}) has produced both UV-luminous candidates at $z\sim 9$--14 \citep{finkelstein2024} and red, apparently massive systems at $z\sim 7$--9 \citep{labbe2023} in excess of many pre-launch expectations. NIRSpec has since spectroscopically confirmed UV-luminous galaxies at $z\sim 10$--14 \citep{curtislake2023,arrabalharo2023,carniani2024}, raising the question of whether stellar feedback operates differently in these environments, and/or whether the same diagnostic tools need to be revised when applied to high-redshift data.

\vspace{0.2cm}

\noindent {\bf Axis IV: Feedback and environment are interdependent.} Metallicity, dust content, density (or compactness), and galaxy-scale properties like the SFR and SFH change the radiative and mechanical energy a stellar population produces and the response of the ISM that absorbs it. Line-driven winds weaken at low metallicity while the ionising output per unit mass typically rises; infrared radiation can be trapped in compact, dusty HII regions until radiation pressure exceeds the ionised gas pressure \citep{olivier2021,barnes2020}, whereas in evolved disk regions the latter dominates; photon leakage and the density into which supernovae explode vary with the same conditions, therefore affecting subsequent feedback coupling efficiencies. The relative importance of feedback mechanisms, escape fractions, or mass-loading factors therefore set, but are also set by, the properties of the environment. It is therefore important to address whether an apparent change in feedback efficiency across different environments (and redshift) is a change in the coupling with the ISM, a change in the stellar output and ISM structure, or both.


\vspace{0.2cm}

The discussion is structured to follow the chain of inference from the sources of feedback to physical interpretation. We restrict our attention to feedback from massive stars specifically — their radiation, winds, and supernovae — and to its consequences for the ISM and galaxy-scale gas cycle. Magnetic fields also influence the structure and dynamics of the ISM (e.g., \cite{beck2013}) and can therefore modify the response to stellar feedback. They are therefore part of the physical context in which feedback operates, even though their contribution is not routinely quantified alongside the thermal, radiation, and mechanical terms discussed here. A comprehensive treatment of magnetic-field regulation and its interplay with stellar feedback is beyond the scope of this review. We also do not review AGN feedback except where it must be disentangled from stellar feedback observationally.

After providing a short refresher on the involved physical mechanisms and quantities and briefly summarising the historical background of feedback studies, the structure of this paper is as follows. Section \ref{sec:obs_diag} examines the principal observational diagnostics of stellar feedback (ionised gas tracers, mechanical feedback signatures, molecular cloud dispersal timescales, galactic outflow indicators, and Lyman continuum escape fractions) with critical attention to the assumptions and systematic uncertainties that go into each. Section \ref{sec:stellar_pops} turns to the stellar population synthesis models that underpin the interpretation of feedback diagnostics and the challenges that arise when stellar populations are resolved (nearby galaxies) versus unresolved (integrated light). Section \ref{sec:highz} examines how the {\it JWST} era is transforming the observational landscape, highlighting the qualitatively new regimes now probed at high redshifts and how existing diagnostic calibrations are placed under stress, and discusses the connection of high-z Lyman continuum escape and feedback. Section \ref{sec:env} examines how metallicity and dust, and then galaxy-scale properties such as stellar mass, star formation rate, and gas content, modulate feedback-related quantities. 
Finally, Section \ref{sec:conclusions} summarises the previous discussions.

\subsection{A brief overview of feedback mechanisms}
Massive stars influence their environment through several (coupled) physical mechanisms, or channels. As mentioned above, for the purpose of this review we focus on radiation, stellar winds, and SNe. Chemical enrichment is discussed only in its role in modifying subsequent feedback (see Section \ref{sec:env}), and we do not discuss protostellar outflows which operate primarily on dense core and early star formation scales and play a more limited role in setting the feedback efficiencies at the larger scales from protoclusters to galactic, nor do we discuss the effect of feedback on protoplanetary disks (see e.g., \citep{winter2022}). What follows is a brief overview to provide the necessary background to stellar feedback physics. A much more detailed physical description of the various feedback mechanisms and their interplay with star cluster formation the interested reader is referred to \citet{krumholz2019}, while a comprehensive discussion about feedback in the context of the formation and lifetimes of giant molecular clouds is given in \citet{chevance2023}. For completeness, we mention cosmic rays but will not discuss these further in the rest of the paper.

\textit{Photoionisation and radiation pressure.} Massive O- and early B-type stars emit the majority of their bolometric luminosity at far- and extreme-UV wavelengths. Photons with $h\nu$ $>$ 13.6 eV ionise hydrogen, heating the surrounding gas to T $\sim$ 10$^{4}$ K and creating an HII region whose thermal overpressure drives expansion into the surrounding neutral and molecular ISM \citep{osterbrock2006}. This process, together with stellar winds, is considered among the primary mechanisms disrupting or dispersing natal gas on timescales of 1-5 Myr prior to the first SN explosions (\citep{dale2014,geen2016,lucas2020}; see Section \ref{sec:clouds}). In addition to thermal pressure from photoionised gas, radiation exerts a direct radiative force on dust grains through absorption and scattering. At high gas surface densities (e.g., in compact HII regions, starbursts, or the early Universe), this radiation pressure on dust can exceed thermal pressure and potentially drive large-scale outflows even without SNe \citep{murray2010,hopkins2012}.

\textit{Stellar winds.} Massive stars lose mass through line-driven stellar winds powered by the radiation field, injecting mass, momentum, and mechanical energy into the surrounding ISM \citep{vink2022}. Additionally, during later evolutionary stages, Wolf–Rayet stars can develop denser winds with terminal velocities of order $10^{3}$ km s$^{-1}$, and although the Wolf–Rayet phase is short, these winds can substantially affect the ISM before core collapse. The wind mechanical luminosities of individual luminous O and Wolf–Rayet stars can reach $L_{\rm w}\sim10^{37}$–$10^{38}$ erg s$^{-1}$ (n.b.~that these are peak values and do not persist throughout the full 3–10 Myr lifetime of massive stars), and winds inflate hot, low-density cavities within HII regions and natal molecular clouds, thereby driving shocks and changing the ISM density and morphology into which subsequent CCSNe explode \citep{dwarkadas2023}. As will be discussed in Section \ref{sec:sps_forward}, corrections to wind properties have lowered O- and Wolf–Rayet-star mass-loss rates, revisions that can reduce the inferred mechanical input by factors of a few (e.g., \cite{bjorklund2021}) with important consequences for the evolution of massive stars and pre-SN feedback.

\textit{Core-collapse supernovae.} The deaths of massive stars as CCSNe release $\sim 3 \times 10^{53}$ erg of energy (predominantly as neutrinos), of which $\sim 10^{51}$ erg couples to the ejecta as kinetic energy \citep{woosley1986}. This kinetic energy drives a blast wave into the surrounding ISM, evolving through an initial free-expansion phase, an energy-conserving Sedov–Taylor phase, a radiative pressure-driven and a momentum-conserving snowplow phases before merging with the ambient medium. A key uncertainty for feedback is the fraction of the canonical 10$^{51}$ erg that thermalises and performs mechanical work before being radiated away. This efficiency depends on the pre-SN ambient density and clearing done by previous feedback mechanisms \citep{gentry2017,lucas2020}. The importance of clustering of supernovae within star-forming regions, that is, repeated explosions inside a common superbubble that maintain a hot, overpressured interior for tens of Myr, increasing the momentum injection efficiency \citep{gentry2020} and ultimately facilitating the launching of large-scale galactic outflows \citep{fielding2018}, is another aspect that is observationally less well constrained (but will not be discussed in this paper; see e.g., \citet{sharma2014}).

\textit{Chemical enrichment.} SN explosions and stellar winds enrich the ISM with nucleosynthetically processed material — oxygen, iron, silicon, carbon, and other metals. This chemical feedback is not energetically dominant but has consequences on the amount and efficiency of feedback produced by the subsequent generations of stars. But while enrichment couples successive generations of stars and their feedback, metallicity acts on feedback in competing directions: higher-metallicity stars drive stronger line-driven winds, and a higher-metallicity ISM contains more dust for radiation pressure to act on, both of which enhance feedback. At the same time, a metal-rich ISM cools more efficiently, increasing radiative losses, and metal-rich massive stars emit a softer ionising spectrum with fewer ionising photons per unit mass. At low metallicity, the ionised gas is consequently hotter \citep{geen2016}. The net effect of enrichment on subsequent feedback, i.e. whether this loop is self-regulating or self-reinforcing, is therefore not established a priori and depends on the feedback channel and environment (see Section \ref{sec:env}).
Further, enrichment of the CGM, observed at metallicities of $\sim$0.1-0.3 Z$_{\odot}$ at large radii \citep{tumlinson2017}, provides evidence that processed material has been transported beyond the star-forming ISM of the disk, partly mediated by feedback via galaxy-scale outflows. 

\textit{Cosmic rays.} Feedback from massive stars also accelerates cosmic ray (CR) particles, with efficiencies of order $\sim$10\% often adopted in models \citep{ruszkowski2023}, although the acceleration efficiency and subsequent coupling remain uncertain. Unlike thermal energy, CRs can survive for tens of Myr in the diffuse ISM. Instead they stream and diffuse through the ISM on long timescales, providing a "slow-release" energy reservoir that can maintain pressure support and drive outflows after the individual supernova remnant has faded \citep{girichidis2022}. CR feedback (and its importance relative to the other stellar feedback mechanisms) in galaxy evolution is reviewed in \citet{ruszkowski2023} and is not discussed further in this paper.

\subsection{A quick history of feedback studies}
The modern understanding of stellar feedback as a dynamical process governing ISM structure and galaxy evolution has its roots in the earliest quantitative descriptions of HII regions and diffuse interstellar gas. Oort \& Spitzer \citep{oort1955} showed that photoionisation by massive stars can drive the acceleration and photoevaporation of surrounding neutral clouds, providing an early quantitative treatment of photoionisation feedback, while similarly early analytical studies showed that photoionisation can produce an overpressured HII region that drives an ionisation front and shock into the surrounding neutral ISM \citep{kahn1954}. Subsequent work developed the classical expansion solution \citep{spitzer1978}. Castor, McCray \& Weaver \citep{castor1975} and Weaver et al. \citep{weaver1977} derived the analytic energy-driven bubble solution for wind-blown cavities in a uniform ISM, while Chevalier \& Clegg \citep{chevalier1985} provided the corresponding solution for a thermally-driven starburst wind from a nuclear region, establishing the foundational model for the launching of kpc-scale outflows.

Important progress was made with the introduction of the three-phase ISM model of McKee \& Ostriker \citep{mckee1977}, who argued that the statistical overlap of SN remnants naturally produces a hot, pervasive, interconnected ISM in approximate pressure equilibrium with the warm and cold phases. This model provided the framework within which subsequent ISM feedback physics is interpreted, and it identified SN feedback as the dominant driver of ISM phase structure. A subsequent important quantitative treatment of stellar feedback in the context of galaxy formation and evolution was provided by Dekel \& Silk \citep{dekel1986}, who argued that supernova-driven winds could efficiently expel gas from dwarf galaxies residing in dark matter halos with characteristic circular velocities below $\sim$100 km s$^{-1}$. Their model demonstrated how feedback could suppress star formation in low-mass galaxies by removing material, and reproduce the observed scaling relations of these systems laying the foundation for modern feedback-regulated models of galaxy formation and evolution.

Observational evidence for starburst-driven galactic winds emerged from studies of extended ionised gas in nearby starburst galaxies with optical spectroscopy revealing large-scale outflows and kinematic signatures consistent with superwinds \citep{mccarthy1987,heckman1990}, later expanded by subsequent systematic observational programmes \citep{lehnert1996}. Large spectroscopic surveys using rest-frame UV absorption lines showed that galactic-scale outflows are ubiquitous in star-forming galaxies and quantified their kinematics, covering fractions, and scaling relations, providing key observational benchmarks for modern galaxy formation models (e.g., \cite{steidel2010,martin2012}).

The field was transformed during the 2010s and 2020s by two parallel developments, at the same time that \citet{bally2011} wrote about the {\it feedback ladder} of increasingly powerful feedback mechanisms. First, the advent of large integral field spectroscopy surveys such as CALIFA, SAMI, and MaNGA \citep{sanchez2012,bryant2015,bundy2015}, together with later high-resolution programmes covering nearby galaxies like NGC~300 \citep{mcleod2020,mcleod2021}, NGC~7793 \citep{dellabruna2021}, M~83 \citep{dellabruna2022}, and surveys such as PHANGS-MUSE \citep{emsellem2022} and SIGNALS \citep{rousseau2019}, which enabled spatially resolved measurements of HII region properties, ISM kinematics, chemical abundances, and ionisation conditions across statistically significant samples of nearby galaxies and the HII regions within them. Second, cosmological hydrodynamical simulations — including large-volume simulations such as EAGLE, IllustrisTNG and SIMBA \citep{schaye2015,pillepich2018,dave2019}, together with high-resolution cosmological zoom-in simulations such as FIRE \citep{hopkins2018} — combined increasingly sophisticated stellar feedback models with sufficient numerical fidelity to make quantitative, testable predictions for galaxy populations across cosmic time. In parallel, high-resolution simulations of the multiphase ISM, such as TIGRESS \citep{kim2017,kim2020}, have resolved the interaction of stellar feedback with the ISM and the launching of galactic winds on smaller spatial scales. These simulations established a framework for interpreting observations in terms of specific feedback prescriptions and their impact on the ISM. The advent of {\it JWST} has now opened qualitatively new observational windows on the high-redshift Universe (Section \ref{sec:highz}), while the forthcoming generation of ELT-class facilities promises spatially resolved studies of galaxies during the peak epoch of galaxy assembly, where many of the uncertainties surrounding stellar feedback are expected to be most acute.

\subsection{Definitions and basic feedback concepts}\label{sec:definitions}

Having defined stellar feedback as the deposition of energy and momentum by radiation, stellar winds, and supernovae into the surrounding ISM and, indirectly, transferred to the CGM (and, further, the IGM), throughout this paper we will at times refer to the feedback mechanisms as "source terms", whereas the fraction of this energy and momentum deposition retained by the ISM, and the resulting effects on it, are referred to as "response". The relevant quantities therefore fall into three categories, i.e.~stellar source terms, local gas responses, and galaxy-scale transport diagnostics, and the principal ones considered here are:

\begin{itemize}

    \item \textit{Ionising photon production rate, $Q(\mathrm{H}^{0})$} 
    (photons s$^{-1}$), i.e.~the intrinsic rate at which a star (or stellar population) emits photons that are capable of ionising hydrogen. Luminosities from recombination lines such as H$\alpha$ trace and constrain the rate of photons absorbed by hydrogen. As will be discussed in Section \ref{sec:halpha}, recovering the intrinsic (injected) $Q(\mathrm{H}^{0})$ requires assumptions about extinction, dust absorption, photon escape, nebular conditions, and the ionising stellar population itself \citep{kennicutt2012}.

    \item \textit{Wind mechanical luminosity, $L_{\rm w}$} 
    (erg s$^{-1}$), corresponding to the rate at which the stellar wind carries kinetic energy away from the star. For a wind with mass-loss rate $\dot{M}_{\rm w}$ and terminal velocity $v_{\infty}$,
    \begin{equation}
        L_{\rm w} = \frac{1}{2}\dot{M}_{\rm w}v_{\infty}^{2}.
    \end{equation}
    Its dynamical effect depends on how much of this energy is retained by the surrounding gas. The remainder may be radiated, transferred across phase boundaries, or advected through a porous shell.

    \item \textit{Supernova mechanical luminosity, $L_{\rm SN}$} 
    (erg s$^{-1}$), corresponding to the explosion energy $E_{\rm SN}$ multiplied by the supernova rate $\dot{N}_{\rm SN}$. For a stellar population, the time-averaged mechanical luminosity is
    \begin{equation}
        L_{\rm SN} = E_{\rm SN}\dot{N}_{\rm SN}.
    \end{equation}
    Two important caveats to keep in mind are that, on the scale of an individual supernova remnant, $L_{\rm SN}$ is not a continuous injection, and the commonly adopted explosion energy $E_{\rm SN}=10^{51}$ erg is the initial value, which can differ from the energy ultimately retained by the ISM after e.g.~radiative losses, mixing, and breakout.

    \item \textit{Feedback-related pressure, $P$} 
    (dyn cm$^{-2}$), often reported as $P/k_{\rm B}$ (K cm$^{-3}$), which quantifies an energy density or momentum flux per unit area and provides a local measure of the capacity of a feedback mechanism to accelerate gas. The pressure terms typically derived are the pressure of the ionised gas, the direct and reprocessed radiation pressures, and the pressure originating from stellar winds. The individual pressure terms are defined below.



    \item \textit{Photon escape fraction, $f_{\rm esc}$}, which measures the fraction of intrinsic photons that escape the region and therefore do not contribute to the ionising budget of an HII region as measured by e.g.~$L$(H$\alpha$),
    \begin{equation}\label{eq:fesc}
        f_{\rm esc} = \frac{Q(\mathrm{H}^{0})-Q(\mathrm{H} \alpha)}{Q(\mathrm{H}^{0})},
    \end{equation}
    where $Q(\mathrm{H}\alpha)$ is the ionising photon rate as inferred from the H$\alpha$ luminosity (as will be described in Section \ref{sec:halpha}). Escape fractions measured on different scales (from an HII region into the diffuse ISM, and from the galactic disk into the CGM) are different quantities, as will be discussed in Section \ref{sec:lyc}. 

    \item \textit{Mass outflow rate, $\dot{M}_{\rm out}$ (M$_{\odot}$ yr$^{-1}$), and mass-loading factor}
    \begin{equation}
        \eta_{M} = \frac{\dot{M}_{\rm out}}{\mathrm{SFR}}.
    \end{equation}
     The mass-loading factor measures the rate of gas transported through a specified surface (e.g., the rate at which mass is transported to larger scale heights due to feedback-driven galactic outflows) relative to the rate at which stars are formed. The mass-loading factor therefore provides a measure for the connection between star formation and stellar feedback in the form of gas removal on galactic scales. Clearly it is not a direct observable, and it depends, for example, on the considered gas phase (molecular, atomic, ionised), the chosen tracer, geometry, the definition of the gas velocity, the characteristic radius from which the outflow is being launched, as will be discussed in Section \ref{sec:outflows}.
\end{itemize}


A prerequisite for stellar feedback studies is the fact that the ISM is multi-phase in nature. Following \citet{mckee1977}, the ISM is broadly characterised by a cold neutral medium (CNM; T $\sim$ 100 K, n $\sim$ 50 cm$^{-3}$), a warm neutral and ionised medium (WNM/WIM; T $\sim$ 10$^{4}$ K, n $\sim$ 0.5 cm$^{-3}$), and a hot ionised medium (HIM; T $>$ 10$^{6}$ K, n $\sim$ 10$^{-3}$ cm$^{-3}$), with molecular clouds (T $\sim$ 10-50 K, n $>$ 10$^{3}$ cm$^{-3}$) forming the cold, dense sites of star formation. Feedback operates across a large dynamic range (from the sub-pc scales of individual feedback-driving massive stars to the kpc scales of galactic disks and the CGM) and acts on the gas phases and involved spatial scales simultaneously, driving the mass exchange between them. This is the main challenge of observational feedback measurements, given that the connection between the feedback-driving stars and star clusters and the multi-scale ISM response requires the phases and scales to be observed at the same spatial resolution. To further complicate matters, different feedback mechanisms couple differently to the different gas phases, and the efficiency with which these mechanisms couple to the ISM likely changes as a function of cosmic time, following the known increase, peak, and decrease in star formation rate across the history of the Universe \citep{madau2014}.

As mentioned in the list above, a framework often used for comparing the relative importance of the various feedback channels (particularly in the context of HII region dynamics and the launching of galactic outflows via supernovae in the pre-cleared ISM) is the measurement of characteristic feedback pressure terms \citep{lopez2014,mcleod2019,barnes2021}. Four pressures are most commonly invoked. The thermal pressure of the ionised gas, 

\begin{equation}
    P_{\rm ion} = 2n_{e}kT_{e}
\end{equation}

arises from the photoionisation heating of the gas to T $\sim$ 10$^{4}$ K by Lyman continuum photons; it is generally a major, and often dominant, contributor to HII region expansion in typical GMC environments and, together with the direct radiation pressure defined next, is the most straightforwardly observable of the four terms, being derivable from the electron density $n_{e}$, typically measured from the [S~{\sc ii}] $\lambda\lambda$6716,6731 doublet, together with an electron temperature $T_{e}$ (either assumed, or better, from auroral-to-nebular line ratios).

The direct radiation pressure, which corresponds to the momentum flux carried by the stellar radiation field, 

\begin{equation}\label{eq:pdir}
    P_{\rm dir} = \frac{L_{\rm bol}}{4\pi r^{2}c}
\end{equation}

where $L_{\rm bol}$ is the bolometric luminosity. If the photons are absorbed or scattered once, the maximum momentum deposition rate is $\dot{p}_{\rm dir}\simeq L_{\rm bol}/c$. This single-scattering limit is expected to be most relevant in compact, high-surface-density star-forming regions, where the radiation flux and dust coupling can be large \citep{murray2010}.

The dust-processed (infrared) radiation pressure,

\begin{equation}
    P_{\rm IR} = \frac{1}{3}u_{\rm IR}    
\end{equation}

as defined by \citet{lopez2014} in their Equation~(6), where $u_{\rm IR}$ is the energy density of the infrared radiation field; it is the dominant pressure term in ultra-compact HII regions in the Milky Way \citep{olivier2021}.

A further radiative term, which is typically not included in observational pressure budgets, is that exerted by resonantly scattered Ly$\alpha$ photons. Roughly two-thirds of recombinations in an HII region produce a Ly$\alpha$ photon. Because of the large line-centre cross-section, these photons can scatter many times in the surrounding neutral gas (line-centre optical depths of $\tau_{0} \sim 10^{6}-10^{8}$ for $N_{HI} \sim 10^{19}-10^{21}$ cm$^{-2}$) before escaping through the line wings. Unlike photons absorbed by dust, Ly$\alpha$ photons are scattered rather than destroyed, and the momentum each one deposits is multiplied by a factor $M_{F}$. Radiation-hydrodynamical simulations find $M_{F} \sim 10-10^{2}$ or more in dust-poor gas \citep{dijkstra2008,smith2017,kimm2018,tomaselli2021}. In these simulations, Ly$\alpha$ pressure can dominate the early, pre-SN momentum budget in metal-poor systems, while dust absorption during the long trapped path reduces $M_{F}$ at higher metallicity. $M_{F}$ depends on the HI column, the gas kinematics and the dust content, none of which are constrained by the emergent Ly$\alpha$ line alone. Further, within the ionised zone of dusty HII regions, Ly$\alpha$ radiation pressure is expected to be minor (e.g., \cite{draine2011}), and the large force multipliers arise from trapping in the surrounding neutral gas.
To our knowledge, the pressure from Ly$\alpha$ photons has not been measured observationally for resolved HII regions, and it is therefore absent from the observational comparisons discussed in this paper. 

Finally, the stellar wind ram pressure, which is commonly parameterised in HII region analyses as the volume-averaged pressure, 

\begin{equation}
    P_{\rm w} = \frac{3\dot{M}_{\rm w}v_{\infty}}{4\pi r^{2}}    
\end{equation}

which differs from the local wind ram pressure by a geometric factor of three and quantifies the momentum flux of the stellar wind at terminal velocity $v_{\infty}\sim$1000-3000 km/s given the  stellar mass loss rate $\dot{M}$; its conversion to thermal pressure via strong shocks heats the interior of wind-blown bubbles to $T \sim 10^{7}$ K (e.g.,~\citep{rosen2014}), and the temperature and density of the resulting X-ray emitting hot gas can then be used to derive $P_{\rm X}=1.9n_{e}kT_{e}$ from the ideal gas law \cite{lopez2014}. 

The relative magnitudes of these four pressures as a function of e.g., HII region radius, galactocentric distance, star cluster properties, and ambient ISM density has been estimated from observations in HII regions in the Milky Way, the Magellanic Clouds, and nearby galaxies \citep{mcleod2021, barnes2021, olivier2021,dellabruna2022,rowland2024}, and constitutes one of the primary observational diagnostics discussed in Section \ref{sec:obs_diag}.

\section{Observational diagnostics of stellar feedback: what are we actually measuring?}\label{sec:obs_diag}
The observational study of stellar feedback is fundamentally a problem of inference. The commonly used observables — recombination line luminosities, emission line ratios, region morphologies, absorption line Doppler shifts, gas outflow velocities, Lyman continuum fluxes — are a convolution of the underlying stellar physics with the geometry, density structure, ionisation state, and kinematic complexity of the surrounding ISM, and is furthermore averaged over the gas response timescale. The feedback-driving massive stars and the ISM response are not observable at the same spatial resolution, across the same wavelength range, and with the same sensitivity required to disentangle them, except in a handful of nearby regions and systems. Therefore, the fundamental limitation is that observations correspond to more or less integrated properties of the stellar–ISM system projected onto a two-dimensional plane on the sky, averaged along the line of sight and over timescales determined (in large part) by the response of the gas. In very simple terms, deriving both the intrinsic "amount" of feedback being injected by the massive stars and the fractional amount of it that translates into an ISM response requires a chain of assumptions that leads from a measurement to an inferred quantity.
This limitation shapes the discussion of this paper. The subsections that follow therefore operate on three distinct levels: the observables (e.g., H$\alpha$ flux, X-ray luminosity, shell radius/velocity, CO intensity); the inferred physical quantities (e.g., $Q(\mathrm{H}^{0})$, density, gas pressure, mass outflow rate); and the model-dependent physical interpretation (e.g., mechanical luminosities, cloud dispersal times, mass-loading factors). We begin with ionised gas, the workhorse tracer of stellar feedback and extragalactic star formation studies, and then move outward in physical scale: mechanical energy stored in wind-blown bubbles and supernova remnants; the time over which natal molecular clouds remain visible; galaxy-scale outflows; and, finally, the propagation of ionising photons from HII regions into the galaxy ISM and CGM.

\subsection{Ionised gas as a tracer of feedback}\label{sec:halpha}
Recombination lines of hydrogen, most commonly H$\alpha$ and H$\beta$ in the optical, are among the primary tracers of the interaction between massive stars and the ionised ISM. In the simplest case, the H$\alpha$ luminosity of an ionisation-bounded nebula can be used to infer the rate of absorbed Lyman-continuum (LyC) photons. Under Case B recombination and for a characteristic electron temperature of $T_e=10^4$ K, the absorbed ionising photon rate is proportional to the H$\alpha$ luminosity, $Q({\rm H}^0) \propto L_{{\mathrm{H}\alpha}}$,
with the proportionality set by the recombination coefficients \citep{osterbrock2006}. Recombination line measurements are also used to quantify dust extinction through the Balmer decrement. Combined with collisionally-excited lines, they can constrain electron densities and temperatures, while spatially resolved narrow-band imaging and integral field spectroscopy of the ionised gas provide region sizes and, at sufficient spectral resolution, gas kinematics and expansion velocities. When these measurements are combined with a census of the ionising stars---or with a population-synthesis model used as a proxy for that census, appropriately calibrated as discussed in Section~\ref{sec:sps_inverse}---they can be used to estimate the fraction of ionising photons absorbed locally, constrain escape fractions, and infer feedback-related pressure terms such as the direct radiation pressure, $P_{\rm dir}$, and the pressure of the ionised gas, $P_{\rm ion}$, and characterise the effects of feedback on the surrounding gas, including the expansion, compression, or dispersal of ionised structures.

The relation between the luminosity of recombination lines and $Q(\mathrm{H}^{0})$ used as a measure of the ionising photon budget of a young stellar population (also invoked when using H$\alpha$ as a SFR tracer or to convert to $L_{\rm bol}$ \citep{kennicutt2012}) involves a chain of assumptions that become increasingly important when H$\alpha$ is used as a quantitative diagnostic of feedback rather than simply an indicator of recent star formation. These include: (a) the nebula is fully ionisation-bounded, so that all LyC photons are absorbed within the region; (b) the recombination coefficients can be adequately described by a characteristic electron temperature and density; (c) the ionising stellar population is sufficiently well sampled by the assumed IMF; (d) the recombination line emission is not significantly contaminated by non-stellar excitation mechanisms such as shocks or AGN activity; (e) the aperture samples a well-defined nebula rather than a mixture of HII region and diffuse ionised gas (DIG) emission; and (f) absorption of LyC photons by dust within the nebula is negligible or is modelled appropriately.

These assumptions affect the physical interpretation of the inferred $Q(\mathrm{H}^{0})$ or $L_{\rm bol}$ in different ways. For example, density-bounded regions allow LyC photons to escape before producing recombinations, such that H$\alpha$ measures only the absorbed component of the photons emitted by the stellar population. In the most strongly leaking systems, neglecting escape of LyC photons can lead to substantial underestimates of the intrinsic $Q(\mathrm{H}^{0})$ \citep{izotov2016,mcleod2019,ramambason2022}. However, in the case that the intrinsic $Q(\mathrm{H}^{0})$ can be independently inferred (for example, from a complete census of massive stars and their spectral types), the difference between the two separate $Q(\mathrm{H}^{0})$ estimates can therefore be used to constrain an HII region escape fraction $f_{\rm esc}$ (Eq.~\ref{eq:fesc}), with the caveat that the interpretation also depends on dust absorption, matching a given nebula with a given stellar population, and resolution.

\begin{figure}
    \centering
    \includegraphics[trim={0cm 3cm 0cm 1.9cm},width=0.6\linewidth]{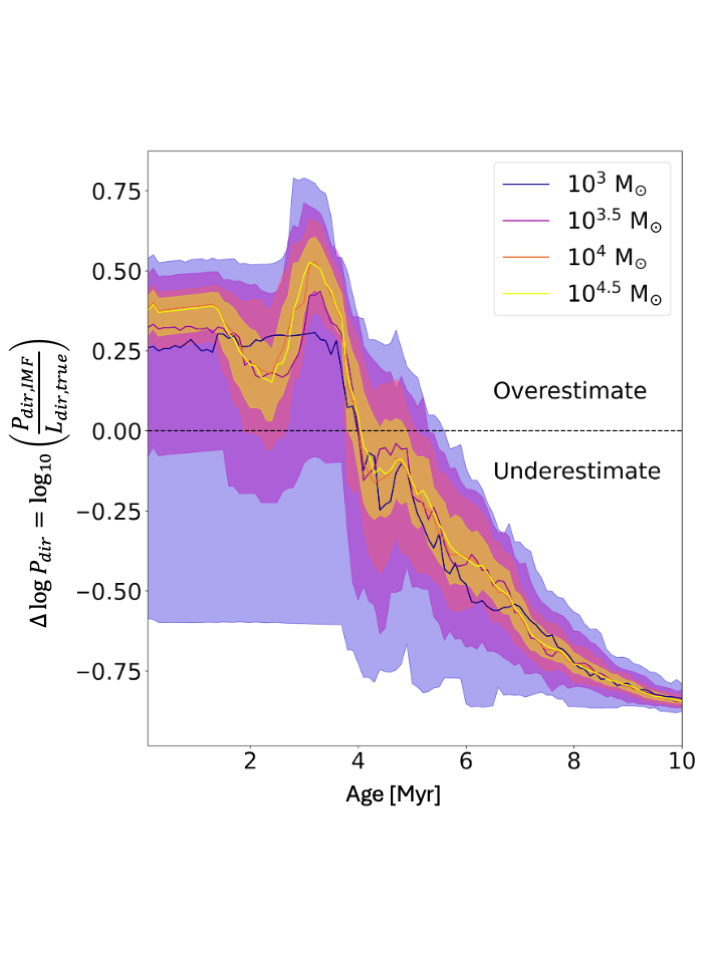}
    \caption{Fractional difference between the direct $P_{\rm dir}$ inferred from the H$\alpha$ luminosity using the fully sampled IMF calibration of \citet{kennicutt2012}, and the intrinsic $P_{\rm dir}$ of stochastically sampled synthetic stellar clusters, as a function of cluster age. The synthetic clusters, generated with the stochastic stellar population synthesis code SLUG \citep{dasilva2012}, span stellar masses of $10^{3}$–$10^{4.5}$ M$_{\odot}$, with 100 realisations at each mass. The increasing scatter (shaded regions) towards lower cluster masses reflects the stochastic sampling of the upper IMF and the resulting variation in ionising luminosity. This figure is adapted from \citet{rowland2024} who plot the fractional difference between bolometric luminosities, but
    at fixed cluster radius, the fractional error in $P_{\rm dir}$ and that in the inferred $L_{\rm bol}$ are equivalent, since $P_{\rm dir}\propto L_{\rm bol}$.}
    \label{fig:rowland}
\end{figure}

At the other extreme, at sufficiently low stellar masses, incomplete sampling of the upper IMF makes the relation between stellar mass and ionising luminosity stochastic. Consequently, converting $L_{\mathrm{H}\alpha}$ to $L_{\rm bol}$ and then to $P_{\rm dir}$ using a fully sampled IMF calibration can introduce age- and mass-dependent scatter and bias. This is illustrated in Figure~\ref{fig:rowland}, adapted from \citet{rowland2024}, who used the stochastic stellar population synthesis code SLUG \citep{dasilva2012} to quantify the fractional difference between $L_{\rm bol}$ inferred from $L_{\mathrm{H}\alpha}$ (using the fully sampled IMF calibration of \citet{kennicutt2012}) and the intrinsic $L_{\rm bol}$ of the stochastically-sampled synthetic clusters. In Figure \ref{fig:rowland} this difference is expressed in terms of $P_{\rm dir}$, since at a fixed region radius $r$, the fractional error in $P_{\rm dir}$ and that in the inferred $L_{\rm bol}$ are equivalent (Eq.~\ref{eq:pdir}) and the stochasticity propagates into the inferred contribution of the direct radiation pressure to the total feedback budget. Their synthetic clusters span stellar masses of $10^{3}$–$10^{4.5}$ M$_{\odot}$, with 100 realisations at each mass. The resulting scatter is largest for low-mass, young populations, while the median bias changes with age in such a way that the fully sampled IMF calibration tends to overestimate $L_{\rm bol}$ ($P_{\rm dir}$) for very young populations and underestimate it at later times \citep{rowland2024}. 

If spatial resolution is low and therefore insufficient to separate individual HII regions from the surrounding diffuse DIG, the measured H$\alpha$ luminosity can include emission that is not powered locally by the associated young stellar population. Across typical spiral galaxies, DIG emission contributes roughly 20--50\% of the total H$\alpha$ emission \citep{haffner2009,mcclymont2024}. In terms of the origin of the potentially contaminating DIG emission, ionising photons leaking from HII regions are generally considered an important source of it \citep{mannucci2021}, but old, low-mass stellar populations can contribute significantly in some systems \citep{belfiore2022}, while shocks and turbulent mixing layers provide additional sources of excitation. Consequently, H$\alpha$ emission integrated over spatial scales of hundreds of pc cannot in general be uniquely associated with the young massive stars responsible for the nearby HII regions, ambiguity that propagates directly into estimates of $Q(\mathrm{H}^{0})$ and, when H$\alpha$ is used to infer $L_{\rm bol}$, into the corresponding estimate of $P_{\rm dir}$.

What this means in practice is that the physical information contained in H$\alpha$ measurements depends on whether the ionising stellar population and the responding gas can be spatially and physically matched, i.e., there is a probable causal connection between the LyC emitting stars and the H$\alpha$ emitting gas. The increasing availability of spatially resolved stellar spectroscopy (from instruments such as VLT/MUSE, Keck/KCWI, CFHT/SITELLE) and high angular resolution imaging ({\it HST} of course but increasingly also {\it JWST}) across entire nearby galaxies has therefore shifted the use of recombination and collisionally-excited lines from a simple luminosity-based tracer towards a diagnostic of the coupling between individual massive stars and star clusters and the surrounding ISM. In particular, three related questions have become accessible: how much of the ionising photon budget is absorbed locally, how the resulting ionised gas responds dynamically, and how both these quantities vary with environment.

The first of these questions can be addressed by comparing the intrinsic ionising output of the stellar population with the recombination line luminosity of the surrounding ISM, folding these into a photon escape fraction (\ref{eq:fesc}). In the LMC star-forming complexes N44 and N180, \citet{mcleod2019} used VLT/MUSE spectroscopy to identify and characterise the individual O-type and WR stars and compared their intrinsic stellar ionising photon budgets, $Q(\mathrm{H}^{0}_{\ast})$, with the Case B rates inferred from the measured H$\alpha$ emission, $Q(\mathrm{H}\alpha)$. The difference between the stellar and nebular estimates implies substantial leakage, with inferred HII region escape fractions of $\sim0.4$--$0.8$. An analogous pilot analysis in the nearby ($D\sim$ 2 Mpc) galaxy NGC~300, combining MUSE $<$10 pc resolution spectroscopy with {\it HST} imaging, identified 13 previously unknown O-type stars and four WR stars in two of the brightest HII region complexes \citep{mcleod2020}. In this case, $Q(\mathrm{H}\alpha)<Q(\mathrm{H}^{0}_{\ast})$ in only two of the five regions for which the stellar populations were sufficiently constrained, corresponding to inferred escape fractions of approximately $0.3$ and $0.5$. It is important to note, however, that a difference between the ionising photon rate as predicted by the stellar population and as measured from H$\alpha$ can only be taken as evidence for LyC leakage after properly accounting for other mechanisms that would lead to photons being removed from the intrinsic, emitted budget, in primis of course dust absorption, but also (in)completeness of the massive stellar population, and spatial or aperture mismatches between the stellar and nebular measurements. Escape fractions are, therefore, indirect measurements inherent to a specific region and the adopted assumptions. 

Beyond distances of 1-2 Mpc, so beyond the distance of NGC~300, individual massive stars can no longer be resolved and measurements are necessarily made for integrated star clusters, and one moves from direct stellar characterisation to population-level inference. At a distance of $\sim$10 Mpc, \citet{teh2023} combined LEGUS cluster photometry and stochastic SLUG population modelling with SIGNALS H$\alpha$ measurements in NGC~628 \citep{rousseau2019} for a statistical comparison between the predicted stellar ionising photon budget and the nebular emission. For the cluster-nebula associations with sufficiently constrained ionising photon posterior distributions (and accounting for dust extinction in the SLUG modeling), the inferred HII region escape fraction was $f_{\rm esc}\simeq0.09$, with larger values in regions with lower H$\alpha$ luminosities. The inferred $f_{\rm esc}$ values are therefore HII region leakage fractions that depend on how well one can match the ionising sources and the nebulae, as well as on the stellar population synthesis model adopted (see also Section \ref{sec:sps_forward}).

The same comparison can now be made statistically across the sample of nearby spirals of the PHANGS survey (\citep{emsellem2022}) by matching MUSE HII regions to their
{\it HST}-identified stellar associations
\citep{scheuermann2023,scheuermann2026}. For regions linked one-to-one
to a single young association (that is, each region has a single, bona fide stellar association that can me matched to it), \citet{scheuermann2026} infer a median HII region escape fraction of $f_{\rm esc}\sim82\%$. This means that most of the stellar LyC budget does not stay in the H II regions, consistent with DIG being fed by leakage. \citet{scheuermann2026} also discuss how, near the median age of $\sim$2~Myr of the stellar associations, even a 1~Myr uncertainty on the derived association age (and this goes back to the discussion around Figure~\ref{fig:rowland}) can change the derived $Q({\rm H}^{0})$ by a factor of two, underlining that inferred escape fraction inherits the uncertainties of the adopted stellar population synthesis model, on top of the uncertainty coming from the one-to-one spatial matching itself. Again, and together with the LMC, NGC~300 and NGC~628 results above, this shows that region-scale $f_{\rm esc}$ can become a feedback-related quantity once one has control over the completeness of the stellar census or the assumptions made for the stellar population models. Values ranging from $\sim$10\% to $\gtrsim$80\% in nearby galaxies should therefore not be averaged into a single extragalactic escape fraction, and instead are an indication for the range of conditions which will influence how much of the stellar input can end up as feedback locally: an HII region escape fraction of $\sim$0.2 means that most ionising photons are still absorbed locally, so H$\alpha$ remains a viable tracers of the stellar population. Values of $\sim0.5-0.6$ imply that the nebula is strongly density-bounded or porous, and that roughly half of the intrinsic $Q({\rm H}^{0})$ does not end up as local recombinations. 

Now, to the second question of how the resulting ionised gas responds dynamically. Spatially resolved HII region observations can also be used to quantify the dynamics of the ionised gas. Using the same MUSE observations of N44 and N180, following \citet{lopez2014}, \citet{mcleod2019} also estimated $P_{\rm ion}$, $P_{\rm w}$, and $P_{\rm dir}$. They found $P_{\rm ion}$ and $P_{\rm w}$ to dominate over $P_{\rm dir}$ by two to three orders of magnitude, and conclude that the expansion of the analysed HII regions is therefore driven by stellar winds and the pressure exerted by the ionised gas, consistent with previous LMC and SMC pressure calculations \citep{lopez2014}. However, the interpretation of the contribution of stellar winds requires caution in the case of \citet{mcleod2019}, because the dynamical estimate of the wind luminosity is obtained by assuming that stellar winds alone account for the observed bubble expansion. It therefore quantifies the wind input required under that assumption, rather than independently measuring the actual pressure of the hot gas from e.g.~X-rays. hence, if $P_{\rm dir}$ also contributes to the expansion, the wind input required to explain the observed bubble sizes and ages would be lower. \citet{mcleod2019} also infer wind luminosities directly from the characterised stellar population within the regions, finding that in most of them (7 out of 11), this value is smaller the wind luminosity inferred from a Weaver-type bubble model.  The relative dynamical contributions of winds and ionised gas therefore remain model-dependent, particularly when energy leakage, radiative losses, or environmental confinement affect the bubble evolution, as will be discussed in the next section.

The subsequent NGC~300 analysis similarly found $P_{\rm ion}\sim P_{\rm w}\gg P_{\rm dir}$ \citep{mcleod2020}. Exploiting the full 35 arcmin$^2$ MUSE coverage of the star-forming disk of NGC~300, \citet{mcleod2021} then extended the analysis to the full population of $\sim$100 covered HII regions and found that the inferred pressure terms vary systematically with galactocentric environment, in particular along the negative metallicity gradient, demonstrating the interdependence of feedback and environment. In fact, when compared to $P_{\rm ion}$ and $P_{\rm dir}$ measurements of Milky Way ultra-compact \citep{olivier2021} and Central Molecular Zone \citep{barnes2020} HII regions, the lower surface density, moderately lower metallicity, and overall more evolved (i.e., larger) HII regions in NGC~300 and the LMC have overall lower pressure terms, adding to the growing evidence that these depend on evolutionary stage and environment. At the scale of individual star-forming complexes, the relative importance of the different feedback channels is also expected to depend on the mass of the feedback-driving cluster and the escape velocity of the region, which, together with compactness (radius) and surface density, set the conditions under which photoionisation, direct radiation pressure, winds, and supernovae can remove or regulate gas \citep{krumholz2019}. 

The PHANGS/MUSE survey \citep{emsellem2022,groves2023} expanded pressure-based feedback studies to $\sim$6000 HII regions in 19 nearby spiral galaxies at $\sim50$--100 pc resolution, using extinction-corrected H$\alpha$ as a tracer of the absorbed ionising photon budget, and computing $P_{\rm ion}$, $P_{\rm dir}$, and $P_{\rm w}$ \citep{barnes2021}. The analysis found the three pressure terms to be comparable across galaxies, with $P_{\rm ion}$ only showing marginal dominance. But while this result needs to be interpreted considering the limited spatial resolution (meaning that the inferred pressures depend on assumptions about their sizes and internal structure), what this shows is that a statistically significantly larger HII region sample does not by itself lead to more insight on feedback pressure, unless the probed systems do not span a wider parameter space in terms of environment (e.g., galaxy stellar mass, metallicity, as will be discussed in Section \ref{sec:env}), at least when it comes to comparisons between galaxies.

However, the picture changes if one looks at systematic environmental trends within individual galaxies, as the galactocentric/metallicity trends of feedback-related pressure terms in NGC~300 have already shown. In M~83, for example, \citet{dellabruna2022} found that $P_{\rm ion}$ exceeds $P_{\rm dir}$ by a factor of $\simeq13$ across the disk, but that the two terms become more comparable in the central starburst, where both are enhanced and the HII regions are under-pressured relative to their surroundings. More recently, incorporating {\it JWST} mid-infrared observations into the PHANGS framework, \citet{pathak2025} found that thermal pressure remains dominant over the radiation pressure terms, although the relative importance of radiation pressure increases when weighting by H$\alpha$ luminosity and towards galaxy centres. 

\begin{figure}
    \centering
    \includegraphics[width=1.0\linewidth]{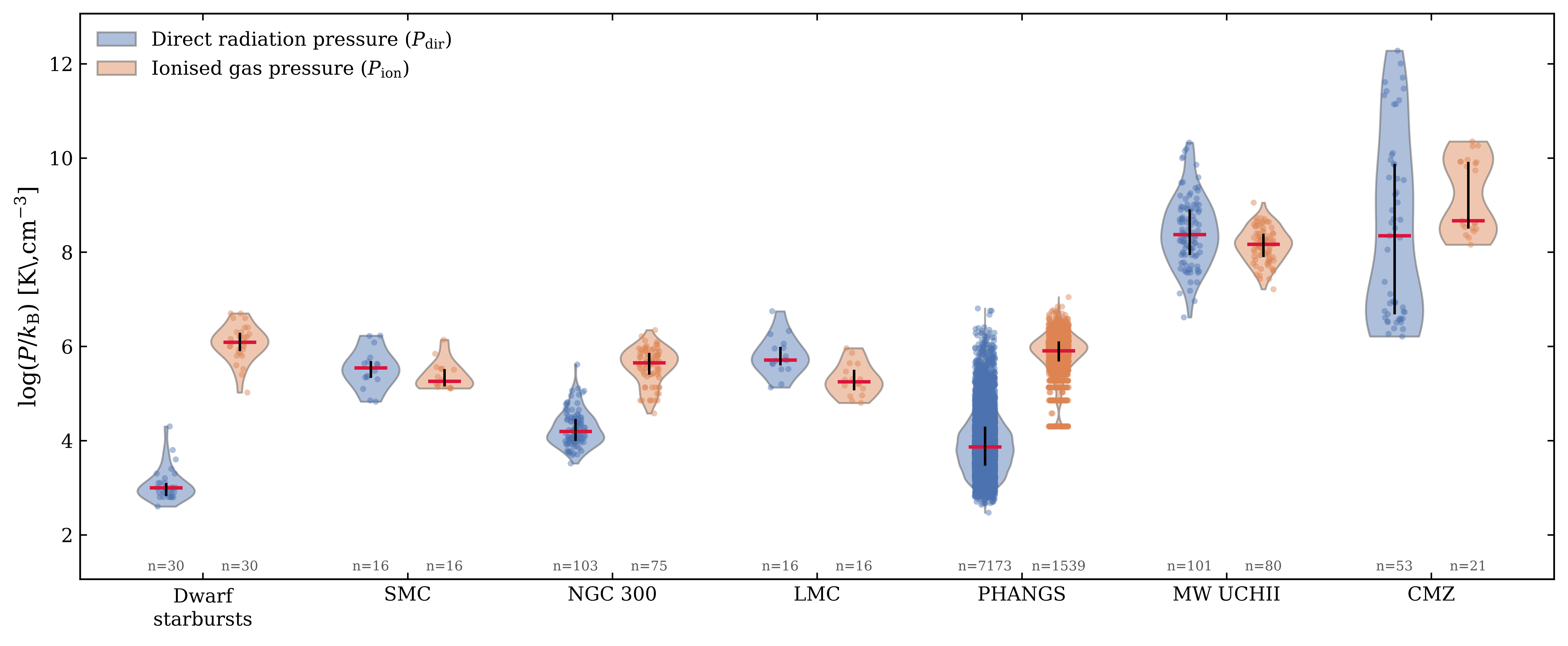}
    \caption{Distributions of direct radiation pressure ($P_{\rm dir}$; blue) and ionised-gas pressure ($P_{\rm ion}$; orange) for HII regions across environments, spanning a range of metallicities and local ISM conditions, from low-metallicity dwarf starbursts and the Magellanic Clouds to high-pressure, compact Galactic environments right: dwarf starbursts \citep{rowland2024}, the SMC \citep{lopez2014}, NGC~300 \citep{mcleod2021}, the LMC \citep{lopez2014}, the PHANGS nearby spiral sample \citep{barnes2021}, Milky Way ultra-compact HII regions \citep{olivier2021}, and the Central Molecular Zone (CMZ; \cite{barnes2020}). Violins show the full distribution for each sample, with individual measurements overplotted as points; red bars mark the median and black bars the interquartile range. Sample sizes ($n$) are indicated beneath each violin. Pressures are shown as $\log(P/k_{\rm B})$ in units of K cm$^{-3}$. Wind ram pressure and X-ray thermal pressure are not plotted; they are different quantities from $P_{\rm dir}$ and $P_{\rm ion}$ and are discussed in Section~\ref{sec:mechanical}.}
    \label{fig:pressures}
\end{figure}

Together, all of the above discussed pressure term results show that the relative importance of different feedback mechanisms depends both on the physical conditions of the surrounding ISM and on the spatial scale of the region being measured. This is further illustrated in Figure~\ref{fig:pressures} where distributions of $P_{\rm dir}$ and $P_{\rm ion}$ measurements for HII regions spanning different types of environments are compared: from dwarf starbursts \citep{rowland2024} and the Magellanic Clouds to the nearby spirals such as NGC~300 and the PHANGS galaxies, but also the already mentioned Milky Way ultra-compact HII regions (UCHIIs), and the Galactic Central Molecular Zone. Outside the densest Galactic environments, $P_{\rm ion}$ typically exceeds $P_{\rm dir}$ by one to several dex, whereas the two become more comparable in UCHIIs and the CMZ, where both can reach $\log(P/k_{\rm B})\gtrsim8$. While one could be tempted to infer a trend of increasing pressure terms with increasing metallicity and ambient density from left to right in Figure~\ref{fig:pressures}, the above discussion cautions us to do so, given the variety of assumptions and measurements that led to these pressure terms. But we will get back to this in Section \ref{sec:env}.


\subsection{Mechanical feedback diagnostics: winds, bubbles, supernova remnants}\label{sec:mechanical}
The ionised gas diagnostics of the previous subsection tell us about photon absorption and about the internal pressure budget of HII regions. And just like radiative feedback quantities, the mechanical energy injected by stellar winds and supernovae is inferred rather than directly observed. 

In the context of this paper, there are three related quantities that need to be distinguished: the momentum flux injected by stellar winds, the thermal pressure of the hot gas, and the fraction of mechanical energy retained or thermalised in the ISM. In the simplest picture, a fast stellar wind (or a supernova blast wave) shocks against the surrounding gas, creating a hot, low-density interior that does $PdV$ work on a swept-up shell. If the injected energy is retained in that interior, the expansion follows a similarity solution. For a continuous mechanical luminosity $L$ into a uniform medium of density $n_0$, \citet{weaver1977} obtained $R\propto n_0^{-1/5}L^{1/5}t^{3/5}$ (the Castor--Weaver or Weaver bubble). An instantaneous supernova energy $E_{\rm SN}$ into a uniform medium yields the related Sedov--Taylor remnant. Observationally, the mechanical energy input from stellar winds and supernovae can be inferred through several diagnostics, including (i) the X-ray luminosity and spectral temperature of hot gas in wind-blown bubbles and supernova remnants; (ii) the expansion velocity, radius, and morphology of ionised or shock-heated shells, measured from emission line kinematics; and (iii) the sizes, expansion rates, and radiative properties of individual supernova remnants. These observables are then interpreted using dynamical and radiative models—for example, Weaver-type energy-driven bubble solutions for wind-blown bubbles and Sedov–Taylor dynamics, together with shock models, for supernova remnants.

Interpreting these observables as a mechanical luminosity requires several assumptions: (a) the hot interior retains a substantial fraction of the injected energy, such that the observed radius, expansion velocity, and age, or the X-ray luminosity and temperature, can be used through a dynamical/radiative bubble model to constrain $L_{\rm mech}$; (b) the ambient medium can be approximated as smooth and uniform, as in the classical Weaver similarity solution \citep{weaver1977}; (c) energy losses through shell leakage, radiative cooling, mass loading, and turbulent mixing are either negligible or adequately described by the adopted model, although these processes can substantially reduce the energy retained by the hot interior \citep{chu2008,lancaster2021}; (d) the momentum flux inferred from the feedback-driving stellar population, $\dot{M}_w v_\infty$, can be directly compared with the thermal pressure of the hot gas, despite representing different stages of the feedback chain; (e) a core-collapse supernova couples $10^{51}\,{\rm erg}$ of kinetic energy to the remnant \citep{thornton1998}. Hence, the inference measures a combination of mechanical power, ambient density, age, \emph{and} energy retention.

A particularly clear illustration of the limitations of the Weaver bubble picture (where the injected energy remains confined, or trapped) comes from observations of young, massive star-forming regions, which consistently show less hot, X-ray emitting gas than expected for such a fully confined Weaver bubble. In the Carina Nebula, \citet{harperclark2009} found that the Castor--Weaver solution overpredicts the observed X-ray luminosity by a factor of $\sim$60, and predicts an expansion velocity about four times larger than what is observed. When the observed radius and age are used to constrain the model, the discrepancy in $L_{\rm X}$ is even larger. These discrepancies point to the fact that the assumption that the injected energy remains confined within a closed, energy-conserving bubble, is very likely incorrect. \citet{harperclark2009} suggested that the energy is advected through a leaky shell, and that the hot gas in this particular region is not dynamically dominant and is approximately in pressure equilibrium with the $10^{4}$ K photoionised gas. Evidence for energy leakage is also found in 30~Doradus, where \citet{lopez2011} measured the X-ray pressure at the bubble rim finding it to be substantially below those expected for a fully trapped wind, consistent with energy escaping through porous channels in the shell. The same observations suggest that real bubbles span a continuum of confinement rather than falling into the idealised trapped-or-free limits. The evolved Magellanic Cloud (LMC and SMC) HII regions studied in \citet{lopez2014} appear to occupy exactly such an intermediate regime between a fully trapped Weaver interior and a freely flowing Chevalier--Clegg wind. Moreover, the X-ray luminosity may simply not be tracing the entire energy budget: \citet{rosen2014} found that the cooling luminosity of the Carina X-ray plasma represents only a small fraction of the injected wind power, implying that most of the mechanical energy is actually not accounted for by the observed X-ray emission. The implication is that X-ray properties alone cannot be used as a measure for stellar wind feedback, and this is corroborated by \citet{dwarkadas2023} who note that observed X-ray temperatures can lie one to two orders of magnitude below the Weaver post-shock temperature. Hence, inferring $L_{\rm w}$ from an idealised Weaver model can introduce large systematic errors.

While the closed, uniform Weaver interpretation might be inadequate at the level of individual systems, a more general problem with quantifying mechanical feedback from the hot, X-ray emitting gas alone does not tell us what happened to the injected mechanical energy that did not result in said X-ray emission. This energy can escape through a porous or ruptured shell, be transferred to cooler phases through turbulent mixing and subsequently radiated, or vent from the galactic disk following breakout. These processes have different consequences for the surrounding ISM and CGM, with the caveat that they can produce similar X-ray observables. For example, \citet{lancaster2021} show that turbulent mixing at the hot/cold interface can remove up to $\sim$99\% of the injected wind energy through radiation at $T\sim10^4$--$10^5\,{\rm K}$, while \citet{fielding2018} find that clustered supernovae can vent a substantial fraction of their energy after breaking out of a the disk. And further, the classical Weaver solution also includes thermal conduction and evaporation at the hot/cold interface, which modify the X-ray-emitting structure of the bubble \citep{weaver1977}. Hence, the same observed $L_X$ deficit can arise from physically different forms of energy loss or redistribution, providing even more evidence for the fact that X-rays might not be the best tracer for mechanical feedback measurements. 

Further, it is important to keep in mind that the different quantities that are taken as a measure of mechanical feedback are not equivalent: the momentum flux from stellar winds, $\dot{M}_w v_\infty$, describes the mechanical input from the stellar population, whereas the thermal pressure inferred from X-ray emission as just discussed, $P_X$, describes the state of one phase of the surrounding gas without accounting for energy losses, but neither of these two quantities measures the {\it total} mechanical energy transferred to the ISM. For example, \citet{lopez2014} and \citet{mcleod2019} find that the pressure from stellar winds can be comparable to, and in some cases exceed, $P_{\rm ion}$ in evolved HII regions in the Magellanic Clouds, and in general find $P_{\rm dir}<P_{\rm w}<P_{\rm ion}$. However, while P$_{\rm w}$ and $P_{X}$ are both attributed to stellar winds, the former is derived from the terminal wind velocities and mass loss rates of the detected massive stars, while the latter is inferred from the hot X-ray gas. Comparing these two quantities for the same regions and stellar populations would provide a direct test of how the wind source term maps onto the hot-gas pressure. But one can conclude that stellar winds are dynamically important, which is supported by smoothed-particle hydrodynamic simulations of star-forming molecular clouds that include prescriptions for both stellar winds and photoionisation \citep{dale2014}. These simulations show that winds can clear gas and create cavities in the ionised component, producing morphologies that are more consistent with observations than those in corresponding simulations without wind feedback.

The observational consequence of the above is that a robust estimate of mechanical feedback requires the various observables to be combined with the information on the stellar population, the ambient density and geometry, the age of the region, and the distribution of gas across different phases, from the cold and warm ISM to the hot, X-ray-emitting component. And, as for other feedback mechanisms, the relevant and unconstrained question again is how much of the injected mechanical energy remains in each gas-phase and on what time- and spatial scales it acts on the ISM. This also pertains to the interpretation of a canonical supernova energy as an observationally calibrated coupling efficiency. A core-collapse supernova is conventionally assigned an explosion energy of $\sim10^{51}\,{\rm erg}$, but the fraction of that energy that remains dynamically relevant to the surrounding ISM again depends on the density and structure of the ambient medium, the degree of pre-SN clearing, radiative losses, SN clustering, and whether the remnant subsequently breaks out of the disk \citep{thornton1998,dale2014,geen2016,lucas2020,fielding2018}. Importantly, whether a supernova explodes into dense natal gas or into a pre-cleared cavity is a separate coupling problem and depends on the timescale over which massive stars clear their natal environment, which connects directly to the molecular-cloud and star formation timescales discussed in the following subsection.


\subsection{Molecular cloud dispersal times}\label{sec:clouds}
The feedback-related pressure terms discussed in the previous sections provide a measure for their relative importance, suggesting that these should be investigated as a function of and across different environments, and introducing the concept of pre-supernova feedback (that is, the relative amount of radiative and mechanical feedback affecting the ISM prior to the explosion of the first SNe). However, the pressure terms do not provide a measure for the timescales of what happens to the ISM in terms of star formation and molecular cloud dispersal once the first stars formed in them start to feed back. Crucially, knowing a "feedback timescale" would provide insight on whether the first supernovae explode into dense gas or into an ISM that has been cleared (or, at the very least, affected) by pre-supernova feedback channels, and therefore how efficiently mechanical (SN) energy couples to the cold ISM (Section~\ref{sec:mechanical}). While following individual clouds through their full lifecycle is observationally difficult, co-spatial multi-wavelength observations enable a statistical view of the lifecycle of GMCs when averaged across entire (nearby) galaxies. With sufficient resolution, maps of different tracers will contain a statistically representative number of star-forming regions so that, under the assumption that they independently cycle through the same evolutionary phases, the relative occurrence of each phase is a measure for the timescale of that phase.

This statistical approach exploits the scale-dependent spatial offset between two tracers of two different phases, for example a tracer of the cold, molecular gas for the pre-stellar phase (such as CO) and a tracer of more or less embedded star formation (e.g., 21 $\mu$m or H$\alpha$, as will be described below). By placing large, kpc-scale apertures on the two tracer maps, the apertures will contain emission peaks in both tracers close to the galaxy average. Within apertures on cloud scales, however, they separate: apertures centred on molecular peaks are gas-rich and star-poor, while apertures centred on H$\alpha$ peaks are the reverse. Plotted as a function of aperture size, that divergence is the characteristic "tuning fork" \citep{schruba2010}. This statistical method converts that scale-dependent spatial decorrelation of star formation tracers into phase durations, under a model of independent regions cycling through the same lifecycle \citep{kruijssen2014,kruijssen2018}, and has been applied to the galaxies of the PHANGS survey using CO and H$\alpha$ \citep{chevance2020,kim2022} and, with {\it JWST}, CO and mid-infrared dust emission \citep{ramambason2026}.

The physical timescales extracted from a tuning fork analysis, such as a CO-bright phase $t_{CO}$, a tracer overlap timescale sometimes termed as feedback time $t_{\rm fb}$, and a cloud-scale star formation efficiency, are therefore timescales and efficiencies associated with the visibility and spatial co-existence of the adopted tracers. Interpreting these quantities as physical GMC lifetimes or feedback-driven cloud dispersal times requires additional assumptions. These include: (a) star-forming regions can be described statistically as independent regions evolving along a common sequence of phases (crudely: gas $\rightarrow$ gas + stars $\rightarrow$ stars); (b) the adopted gas tracer reliably traces the molecular cloud phase (e.g., H$\alpha$ traces the relevant young massive star/HII region phase, 21 $\mu$m traces the embedded star formation phase), (c) the system is sufficiently close to statistical steady state; and (d) the spatial distribution of independent regions is adequately described by the assumed statistics and the characteristic separation scale is resolved. In addition, the conversion of the fitted relative timescales into absolute times requires an externally calibrated reference timescale.

An important caveat concerns the interpretation of $t_{CO}$ as an H$_{2}$-cloud lifetime, because CO is not a direct tracer of the total reservoir H$_{2}$, and the disappearance of CO does not imply the disappearance of H$_{2}$. In regions with low metallicity, low shielding, or strong UV radiation fields, H$_{2}$ can survive where carbon is predominantly C I/C II rather than CO, producing CO-dark molecular gas \citep{chevance2023}. CO-based tuning-fork measurements may therefore measure the lifetime of the CO-bright phase, which may not coincide with the lifetime of the underlying molecular structure. More generally, the disappearance of CO–H$\alpha$ spatial correspondence need not uniquely imply physical cloud dispersal; relative drift between molecular clouds and young stars, for example, can also lead to a tuning fork. Progress on this particular point would require implementing a metallicity-dependent $\alpha_{\rm CO}$ (the factor converting CO luminosity to molecular hydrogen mass and that currently enters the timescale measurements as a systematic uncertainty) as a measured function.

It is perhaps more useful to think of the timeline as nested tracer visibilities. That is, how long a region remains detectable in the chosen gas tracer before it becomes predominantly associated with the stellar tracer. As depicted in \citet{ramambason2026}, a simplistic description of the sequence is: CO-bright gas becomes visible before (massive) stars form; an obscured overlap follows, in which CO and 21~$\mu$m coexist but H$\alpha$ does not (the stars are still embedded but already producing feedback); an exposed overlap follows, in which CO, 21~$\mu$m, and H$\alpha$ are seen together; and isolated H$\alpha$ or 21~$\mu$m remains after CO has disappeared from the beam. \citet{ramambason2026} define the embedded feedback time $t_{{\rm fb},21 \mu m}$ as the CO--21~$\mu$m overlap regardless of H$\alpha$, and the obscured time as the difference $t_{\rm obscured} = t_{{\rm fb},21 \mu m} - t_{{\rm fb}, \mathrm{H}\alpha}$. 

The comparison that then matters for pre-SN versus SN feedback is whether this overlap is shorter or longer than the time it takes the first core-collapse supernovae (CCSNe) to explode, typically $\sim$3-4~Myr for a standard, instantaneous stellar population \citep{chevance2020, leitherer2014}\footnote{We note that the onset of the first CCSNe of course depends on the adopted stellar evolution model. Moreover, not every massive star necessarily produces a luminous CCSN (e.g., \citet{sukhbold2016}), so that the translation from progenitor mass to explosion energy is not a 1-to-1 mapping. This latter point does not, however, affect the CCSN timing argument. See also the discussion of mass loss and explodability in Section \ref{sec:sps_forward}}. Across 54 PHANGS galaxies, after excluding twelve objects for which only upper limits can be measured, \citet{kim2022} find that the CO--H$\alpha$ overlap is $t_{\rm fb}=1.3$--$5.4$~Myr (with mean of about 3.2~Myr). In a previous study, \citet{chevance2020} had obtained a similar 1--5~Myr overlap in an initial subset of nine of those galaxies, and compared it to the mentioned SN delay time of 4~Myr. This comparison was used as an observational argument for pre-SN dispersal of CO-bright gas, given that in most of the sample the overlap is shorter than 4 Myr. 

However, using H$\alpha$ alone as a tracer for star formation is insufficient. Earlier work using \textit{Spitzer}/24~$\mu$m in a handful of galaxies \citep{kim2021} suggested that including the embedded phase lengthens the overlap by 1--4~Myr, and \citet{ramambason2026} provide the statistical {\it JWST} measurement of that missing piece using MIRI 21~$\mu$m versus CO. Among galaxies that are not blending-limited, the median obscured time is $\sim$0.8~Myr and the median CO--21~$\mu$m overlap is $\sim$3.4~Myr. The obscured phase is $\le 4$~Myr in all 37 analysed galaxies and $\le 1$~Myr in 28 of 37. Only 5 of the 37 galaxies (all massive, high-metallicity barred spirals) have a clearly resolved non-zero obscured phase. 

The observed few-Myr overlap therefore favours pre-SN feedback channels for the disruption of CO-bright gas relative to the $\sim$3-4~Myr delay time of the first CCSNe. However, this still does not isolate which one of the pre-SN feedback channels is the dominant one. Moreover, tuning fork-like analyses have not yet been applied to galaxies with $M_\ast\ll 10^9\,M_\odot$ which would provide an extension to the low-mass, low-metallicity regime. Lastly, galactic dynamics (which contribute to setting molecular lifetimes) are encoded in the inferred timescale through the duration of the CO-visible phase: at low mean molecular surface density $t_{\rm CO}$ follows internal crossing times, while above $\Sigma_{\rm H_2}\sim 8\,M_\odot\,{\rm pc^{-2}}$ large-scale collapse and shear become competitive \citep{kim2022}. 

\begin{figure}[H]
    \centering
    \includegraphics[scale=0.65]{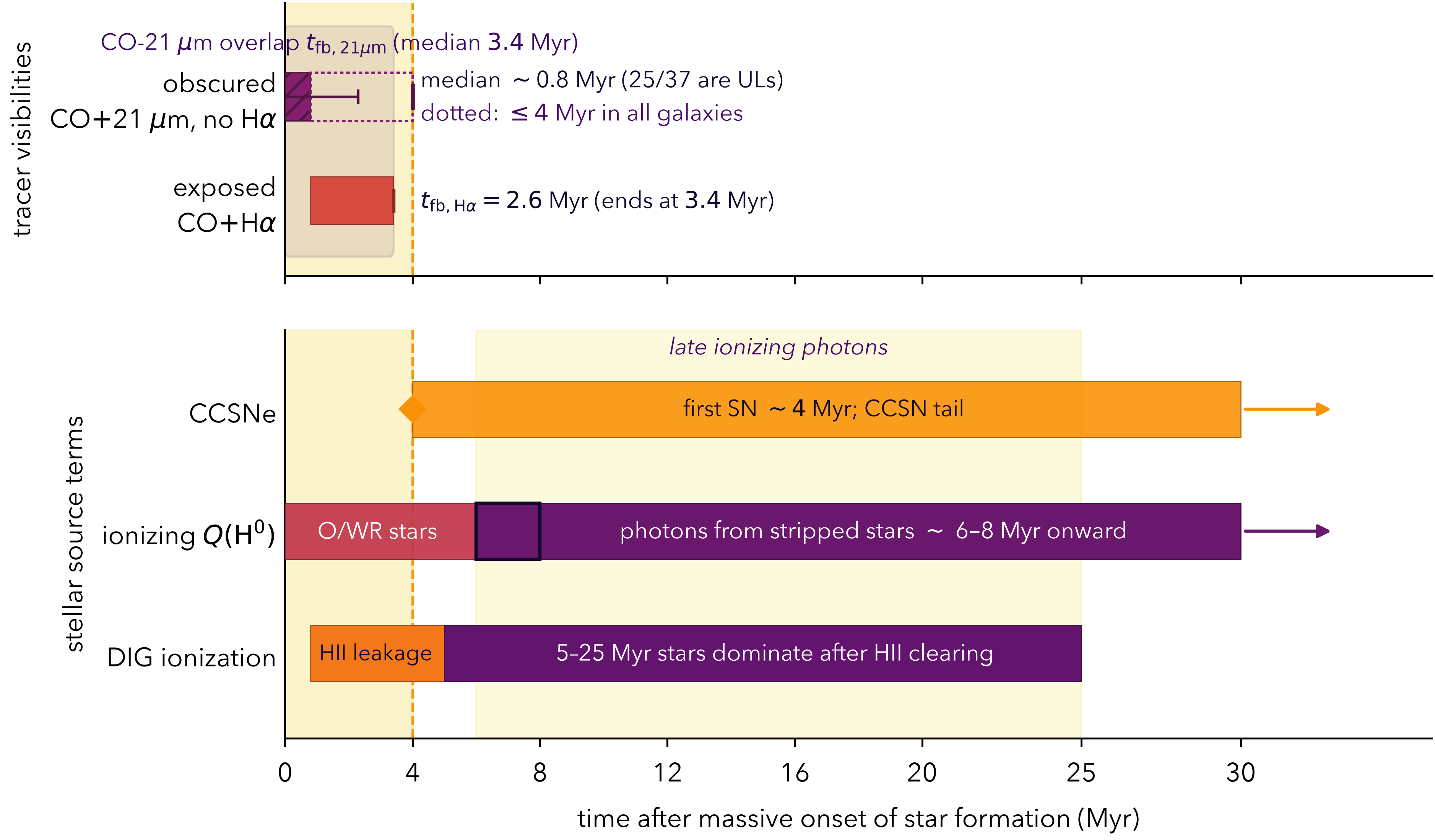}
    \caption{Nested tracer visibilities and stellar source terms after the onset of massive star formation ($t=0$). Isolated CO before $t=0$ is omitted. The upper panel shows PHANGS "tuning fork" timescales of a single GMC--HII cycle, where by definition $t_{\rm obscured}=t_{{\rm fb}, 21\mu m}-t_{{\rm fb},\mathrm{H}\alpha}$ (\citep{ramambason2026}). The hatched bar is the median H$\alpha$-invisible CO--21~$\mu$m overlap, $t_{\rm obscured}\sim$0.8~Myr. 25 of 37 measurements are upper limits (ULs), and the phase is $\le 4$~Myr in all 37 galaxies \citep{ramambason2026}. The exposed bar starts after that median and has width $t_{{\rm fb},\mathrm{H}\alpha}\sim2.6$~Myr, and the parent overlap is $t_{{\rm fb}, 21\mu m}\sim$3.4~Myr. The lower panel shows source terms that run in parallel: the diamond marks the typical minimum delay to the first core-collapse supernova, and the bar is the subsequent CCSN tail; ionisation begins at $t=0$, the contribution of binary star to $Q({\rm H}^{0})$ becomes distinct around 6--8~Myr \citep{eldridge2017,scheuermann2026} and can persist for tens of Myr \citep{gotberg2019}; DIG ionisation begins once HII regions leak and is dominated by stars aged 5--25~Myr after those regions clear \citep{mcclymont2024}. This schematic illustrates tracer visibility and stellar-feedback timescales and is not intended as a deterministic evolutionary track for an individual GMC, nor does it imply that GMCs have been destroyed by the end of the timeline.}
    \label{fig:clocks}
\end{figure}

As the first CCSNe explode around the already mentioned canonical $\sim$4 Myr, CCSN feedback continues for a few tens of Myr as lower-mass progenitors explode and is temporally superimposed with two sources of late ionising photons. Stripped stars emerging from binary envelope evolution produces hot helium stars that keep $Q({\rm H}^{0})$ high after the single-star ionising output has declined, becoming distinct around 6–8 Myr and persisting for tens of Myr \citep{eldridge2017,scheuermann2026,gotberg2023}. Separately, once early feedback has opened HII regions at $\sim$5 Myr, stars aged 5–25 Myr start to dominate the ionisation of the DIG \citep{mcclymont2024}. Stellar feedback therefore continues to shape the ionised ISM well after the first supernovae through late photons in an already-cleared geometry. The sequence of timescales is schematically shown in Figure~\ref{fig:clocks}, where the upper timescale are tuning fork conversions of maps of different tracers, while the lower timescales illustrate which feedback mechanisms continue to operate after the first CCSNe events.

As a final remark on the topic of feedback and molecular cloud dispersal times, we note that the discussion here is heavily biased towards the tuning fork analysis, primarily performed on PHANGS galaxies (which, as already mentioned, span a rather limited range of probed environments). Clearly, other GMC lifetime estimates exist in the literature, as discussed in \citet{schinnerer2024}, and more extreme or diverse galactic regimes need to be included for a more complete and possibly predictive picture. An additional complication is that feedback does not necessarily terminate star formation by directly unbinding the dense molecular gas. Stellar winds may instead suppress continued accretion onto a star-forming region, while supernovae can preferentially remove lower-density gas and prevent star formation from restarting after the dense phase has been disrupted \citep{krumholz2019}. However, it is fair to say that, at the time of writing, the measured few-Myr overlap between tracers already provides strong evidence that the CO-bright phase usually ends before the first CCSNe in the nearby star-forming galaxies studied, that it does not distinguish physical cloud destruction from tracer conversion, drift, or the specific pre-SN mechanism, nor how and if GMC lifetimes depend on environmental properties, which will be discussed in Section \ref{sec:env}.

If natal clouds are cleared on a few-Myr timescale, and if wind-blown bubbles leak or break out of the disk (Section~\ref{sec:mechanical}), gas can leave the star-forming layer altogether. That galaxy-scale consequence is an outflow, and it is the diagnostic that is supposed to connect massive-star physics to the baryon cycle, which will be treated in the next subsection.


\subsection{Galaxy outflow diagnostics: the multi-phase problem}\label{sec:outflows}
Galactic outflows are the macroscopic result of stellar feedback acting on
the ISM on kpc scales. The quantity most commonly used to
characterise an outflow is the already introduced mass-loading factor,
$\eta$, which quantifies the rate of outflowing mass per unit SFR. If $\eta$ is of order unity, star formation and mass ejection are comparable. And while $\eta\gg1$ indicates a high mass flux relative to the SFR, it does not imply that the outflowing gas will be able to escape the galaxy (or reach the CGM). The complicating factor in terms of outflow diagnostics is that outflows are
multi-phase, comprising molecular, neutral atomic, warm ionised, and hot plasma
components. Each phase couples differently to different feedback channels (so that the question of which phase dominates the outflowing mass arises naturally), and each phase requires different tracers with
different assumptions to convert the observable into a mass
\citep{veilleux2005,fluetsch2021}. A measurement of $\eta$ in one phase is
therefore not a measurement of the total mass flux, and does not automatically put constraints on the other, unmeasured phases.

Outflows are commonly identified through gas that is kinematically offset from the systemic star-forming disk component, often producing blueshifted absorption or broad emission line components. In UV absorption, blueshifted low-ionisation metal lines such as Si~{\sc ii}, C~{\sc ii}, and Mg~{\sc ii} trace cool, predominantly neutral or weakly ionised outflowing gas, while higher-ionisation species such as C~{\sc iv} probe more highly ionised gas, including transition or interface material between cooler and hotter phases. Optical Na~{\sc i}~D absorption provides a complementary probe of cool, predominantly neutral gas, particularly in dusty outflows. Conversely, emission line observations decompose the line into a broad (outflow) component superposed on the narrower disk emission, and increasingly
sensitive integral field spectroscopy (IFS) allows these components to be mapped
spatially across nearby galaxies (e.g., \cite{reichardt2025}). In all cases,
the basic observables are a luminosity or absorption column, a velocity, and
some measure of spatial extent, and the inference consists of converting the observables into a mass (and mass outflow rate) and a SFR (see Section \ref{sec:halpha}).

For a given phase, the mass outflow rate is usually estimated as
\begin{equation}
  \dot{M}_{\rm out}\sim \frac{M_{\rm out}v_{\rm out}}{R_{\rm out}},
\end{equation}

where each of these terms introduces phase-specific assumptions. For ionised gas, the mass
can be inferred from the luminosity of a recombination line, provided that
the electron density, temperature, extinction, and ionisation state are
specified. For example, under Case~B recombination,

\begin{equation}
  \dot{M}_{\rm ion}=
  \frac{1.36\,m_{\rm H}\,v_{\rm out}}
  {\gamma_{{\rm H}\beta}\,n_e\,R_{\rm out}}\,
  L_{{\rm H}\beta,{\rm broad}},
\end{equation}

where $L_{{\rm H}\beta,{\rm broad}}$ is the extinction-corrected luminosity
of the broad (outflow) component, $\gamma_{{\rm H}\beta}$ is the Case~B
emissivity at the adopted electron temperature, $n_e$ is the electron
density, and $R_{\rm out}$ is a characteristic radius
\citep{osterbrock2006}. Note the dependence on
$n_e$, which can introduce significant systematic uncertainties given that it is often assumed (e.g., 100 cm$^{-3}$) or otherwise measured from the optical [S {\sc ii}] doublet. Similarly, $R_{\rm out}$ and the geometry of the flow are generally not measured
directly (e.g.~in unresolved systems), and the velocity of the broad component is a tracer-dependent characterisation of what is, potentially, a broad velocity distribution (this latter point is where spatially resolved IFS outflow mapping will certainly deliver constraints for unresolved observations in the upcoming years, e.g.~\citep{reichardt2025}).

Absorption line measurements involve a different set of assumptions compared to emission lines. A general expression is

\begin{equation}
  \dot{M}_{\rm abs} =
  \Omega\,C_f\,\mu m_{\rm p}\,N_{\rm H}\,r\,v,
\end{equation}

where $N_{\rm H}$ is inferred from the observed ionic column through an
ionisation correction and an assumed abundance, $C_f$ is the covering
fraction, $\Omega$ describes the adopted solid angle of the flow, and $r$ is
the characteristic radius of the absorbing material
\citep{heckman2015,fluetsch2021}. Here the dominant uncertainties are
therefore the ionisation corrections, the adopted metallicity,
covering fraction, and the geometry and location of the absorbing gas. These all
enter the conversion from column density to mass flux, and the main consequence is that emission and absorption measurements do not necessarily yield the same $\dot{M}$, even if they probe physically related gas. The same issue applies to the colder phases: atomic gas can be traced via lines such as [C~{\sc ii}]~$158\,\mu{\rm m}$, but converting to a mass requires assumptions about the C$^+$ abundance and excitation. Molecular outflows are commonly traced through CO, with the inferred mass depending on the already mentioned adopted CO-to-H$_2$ conversion factor \citep{bolatto2013,fluetsch2021}, on the excitation, and on the optical depth, and one might not be able to readily use the same assumptions for the outflowing material as for the molecular gas of the galaxy disk.

The SFR in the denominator of $\eta$ is usually obtained by applying a Kennicutt--Evans calibration \citep{kennicutt2012} to an extinction-corrected H$\alpha$ luminosity, to a
FUV continuum luminosity, to a total IR luminosity $L_{\rm TIR}$ or a mid-infrared monochromatic luminosity, or to a hybrid indicator (FUV+$L_{\rm TIR}$,H$\alpha$+$24\,\mu{\rm m}$) that sums the unobscured and dust-reprocessed radiative output from massive stars. These indicators probe different timescales:
recombination emission is dominated by the most recent $\lesssim 5$--$10$ Myr
of star formation, whereas the FUV continuum averages over substantially
longer timescales, and infrared emission can include a contribution from
older stellar populations. Moreover, the conversion assumes a fully sampled IMF and a continuous star formation over the tracer's visibility window, as well as an SPS library (commonly Starburst99 \citep{leitherer1999}). The resulting $\eta$ therefore compares an instantaneous mass flux to a time-averaged and calibration-dependent star formation rate.

These systematics become especially important when the different gas phases
are compared in the same galaxies (as opposed to the same gas phase across different galaxies). For example, in local (U)LIRGs where all three phases have been measured, the molecular phase accounts for $\gtrsim 60\%$ of $\dot{M}_{\rm out}$ in most systems (with the fraction being larger in AGN-hosting galaxies), the Na~{\sc i} phase for $\sim 10\%$, and the ionised phase for a negligible share of the total \citep{fluetsch2021}. In the star-forming subset of these systems, the molecular fraction is lower, and the ionised and atomic gas are often co-spatial and have similar kinematics. In NGC~253, optically thin CO filaments give a conservative molecular $\eta\sim 1$--$3$ \citep{bolatto2013}, comparable to the values of the ionised phase mentioned above. While one phase might dominate over the others, it is important to keep in mind that the three can be part of the same flow, and that a hot, volume-filling wind is able to provide the pressure and ram force that accelerates cooler material, while most of the mass can be contained in the molecular and atomic components. And, as expected, the relative mass in each phase can be environmentally dependent meaning it can vary with galaxy type and outflowing radius.

Once the systematics have been taken into account and $\eta$ is known, the next question then is how the different phases are connected, and what specific feedback mechanism(s) accelerated the gas. Spatially resolved spectroscopy provides one route to answering this question by measuring the velocity, surface brightness, and spatial extent of the outflowing gas, while multi-wavelength observations can determine whether ionised, atomic, molecular, and hot components are co-spatial and kinematically coupled. For example, resolved ionised gas observations can constrain how the outflowing mass flux varies with local star formation intensity. This was done for a sample of dwarf starburst galaxies imaged with the KCWI integral field instrument on Keck, where such a comparison shows that (although the inferred value depends strongly on the adopted extinction correction and other assumptions) the outflows strongly correlate with local SFR surface density, implying that the spatial distribution of the underlying starbursts modulates the spatial distribution of outflow properties \citep{reichardt2025}. 

In terms of what the acceleration mechanism is, one test is to compare the slope of $v_{\rm out}$--$\Sigma_{\rm SFR}$ to a shallow mechanical-energy scaling
($\propto\Sigma_{\rm SFR}^{0.1}$) and to a radiation-momentum scaling ($\propto\Sigma_{\rm SFR}^{2}$). Another test is $\eta$ versus $v_{\rm out}$, compared to the constant-momentum and constant-energy scalings ($\eta\propto v_{\rm out}^{-1}$ and $v_{\rm out}^{-2}$) used in some semi-analytic recipes. The note of caution here is, of course, that these are idealised scaling expectations rather than definite signatures of individual feedback channels. Following such a comparison, \citet{reichardt2025} find slopes of $0.12$--$0.20$ in the dwarf starburst galaxies in their analyses, consistent with a supernova-driven regime rather than the simple radiation-pressure scaling, well below $\propto\Sigma_{\rm SFR}^{2}$. \citet{heckman2015} find no $\eta\propto v_{\rm out}^{-1}$ or $v_{\rm out}^{-2}$ in UV absorption, and integrate a momentum-driven equation of motion for the absorbing clouds (that is, ram pressure of a hot flow, plus gravity). When put into context of the previous subsections where we discussed how pre-SN feedback can clear clouds prior to the onset of SN feedback, outflow measurements lack the timescale aspect and while SNe appear as the primary driver of galaxy outflows, the role and relative importance of pre-SN feedback channels in enabling SN-driven outflows is obvious, but only partially constrained. The same stellar feedback event can deposit energy and momentum into different phases, while subsequent shocks, radiative cooling, mixing, and entrainment
can transfer mass and momentum between them, and a shallow velocity--$\Sigma_{\rm
SFR}$ relation in warm ionised gas, for example, does not make a direct causal argument that the outflow is mechanically (SN) driven. Clearly, this is yet another argument for the fact that observational feedback studies require "the drivers and the driven" to be observed at the same spatial scales across all tracers and phases, so that phase-resolved, channel-distinguishing and timescale-quantifying measurements can be performed in the same systems.




\subsection{Lyman continuum escape and feedback-cleared channels}\label{sec:lyc}
Stellar feedback disrupts and disperses CO-bright natal gas on few-Myr timescales (Section~\ref{sec:clouds}) and can contribute to the formation of low-column-density channels through which radiation and outflowing gas escape. In the scenario that every photon emitted by a massive stellar population is absorbed in the natal HII region, the nebula is ionisation-bounded and the fraction of photons that escape, $f_{\rm esc}$, is zero. A non-zero escape fraction (central to cosmic reionisation arguments, where the escape fraction typically corresponds to a quantity that is averaged over an entire galaxy) requires optically thin sightlines through the neutral/dusty material surrounding the ionising source. Density-bounded HII regions are one limiting configuration that permit this. 

For the purpose of the following discussion, we must distinguish three escape fractions depending on the scale that is being considered: $f_{\rm esc,HII}$ (from HII region scales), $f_{\rm esc,ISM}$ (from the star-forming ISM on galaxy scales), and $f_{\rm esc,gal}$ (beyond the galaxy/halo). We already discussed $f_{\rm esc,HII}$ in the context of using ionised gas emission as a tracer of stellar feedback and, therefore, the fraction of photons that escape and do not enter the local feedback budget, in Section \ref{sec:halpha}. Here, we want to now focus on feedback-related measurements in the context of unresolved, higher redshift systems, moving closer not only to cosmic noon, but also to the much higher redshift galaxies in the very early Universe recently revealed by {\it JWST} that will be discussed in Sections \ref{sec:stellar_pops} and \ref{sec:highz}. From quantifying how many photons have escaped the regions instead of contributing to the local feedback budget, the question becomes where in the galaxy--CGM--IGM system those photons are absorbed. At larger scales, these photons remain part of the feedback budget, in the sense that they have transferred their energy and momentum deposition from the local natal cloud to the surrounding ISM or CGM. The location of the energy deposition has simply moved to larger spatial scales. Observationally, the challenge is therefore to determine how much of the intrinsic ionising photon budget is absorbed locally, how much escapes the star-forming ISM, and how much ultimately reaches the CGM or IGM.

While recombination lines can also be used for larger-scale escape fraction studies, measurements of $f_{\rm esc,ISM}$ and $f_{\rm esc,gal}$ are typically performed in the UV. If H{\sc i} along a sightline is optically thick, the stellar Lyman continuum is absorbed and the observed flux below the Lyman limit at 912~\AA\ is zero. One then converts the residual flux (that is, the leftover ionising continuum after that absorption), or the extinction-corrected H$\alpha$ or H$\beta$ luminosity (which yield the ionising photons absorbed in the nebula under Case~B recombination) into $f_{\rm esc}$ by dividing by the ionising continuum the stars are inferred to have produced, where the latter is inferred from a resolved stellar population, from an extinction-corrected recombination-line luminosity, or from a population synthesis fit to the observed SED or non-ionising UV continuum, at the radius of the measurement. Therefore, the denominator folds an age, star formation history, metallicity, and binary prescription into the $f_{esc}$ calculation. 

At $z=0.2$--$0.4$, individual HII regions are generally unresolved by the apertures used for direct LyC spectroscopy. The Lyman limit falls in the \textit{HST}/COS bandpass, and the aperture encompasses scales of several kpc, so the measured residual LyC flux is already a measurement on a substantially larger scale than the individual nebular regions considered in Section~\ref{sec:halpha}. The Low-redshift Lyman Continuum Survey (LzLCS) provides the largest direct sample in this regime: \citet{flury2022a} detect LyC emission in 35 of 66 galaxies and infer escape fractions ranging up to $\sim50\%$. The important result for the present discussion is the large object-to-object variation: the strongest leakers are compact and often have high $O_{32}$ (i.e., a high degree of ionisation), but many equally compact, high-$O_{32}$ galaxies remain optically thick along the COS sightline \citep{flury2022b}. These observations show that ionising photons can leave the star-forming layer on kiloparsec scales in the presence of channels along particular sightlines.

The connection to stellar feedback becomes more explicit when the stellar populations and geometry of the neutral gas that is used as the tracer are considered alongisde one another. Using a larger set of LzLCS spectra, \citet{flury2025} find that mechanical feedback associated with $\sim$8--10~Myr stellar populations is enhanced in LyC leakers and is associated with the anisotropic gas distributions required for escape, while the highest escape fractions additionally require very young populations that provide the contemporaneous ionising luminosity. Their interpretation is therefore a temporal sequence: mechanical feedback creates or enlarges low-column-density channels, while subsequent young massive stars are the sources of the ionising photons that escape through them. The result connects the feedback timescales discussed in Sections~\ref{sec:clouds} and~\ref{sec:mechanical} directly to the emergent LyC budget. It also illustrates why $f_{\rm esc}$ cannot be interpreted as a simple measure of feedback strength, because a large escape fraction requires both a sufficiently cleared geometry and a sufficiently luminous young ionising population.

At $z\sim3$, the Lyman limit is shifted into the observed-frame optical, allowing direct searches for residual LyC emission from the ground. The measurement is nevertheless fundamentally different from the low-redshift case because the observed LyC has passed through the CGM {\it and} IGM before reaching the observer. The intrinsic escape fraction must therefore be inferred jointly with the transmission probability of intervening H{\sc i}. \citet{shapley2016}, for example, report a clean LyC detection in Q1549-C25 at $z=3.15$ and infer $f_{\rm esc}\geq0.51$ at 95\% confidence after modelling absorption along the line of sight. At still higher redshift, the increasing opacity of the intervening IGM makes direct LyC measurements progressively more difficult and ultimately impractical during the epoch of reionisation. The observational problem then changes from measuring the escaping photons directly to inferring the conditions that permit their escape. Ly$\alpha$ emission, low-ionisation absorption, nebular line ratios, and other tracers of ionisation and neutral gas covering are therefore used as indirect diagnostics of LyC escape \citep{jaskot2014,flury2022b}, but the loss of spatial resolution and the inability to observe the escaping continuum directly make the inference correspondingly more model dependent. 

The stellar denominator in deriving $f_{\rm esc}$ introduces a second, independent uncertainty at high redshift. The intrinsic ionising photon production rate depends on the assumed stellar population model, including its treatment of binaries, stellar rotation, metallicity, and the recent star-formation history. This is the topic of Section \ref{sec:stellar_pops}. Binary evolution can maintain a harder ionising spectrum at ages when single-star populations would predict a much more rapid decline in $Q(\mathrm{H}^{0})$ \citep{eldridge2017}, and simulations find that this can increase the predicted time-averaged LyC escape fraction \citep{rosdahl2018}. This is particularly relevant when feedback and escape are time-dependent: mechanical clearing and the production of ionising photons don't necessarily peak simultaneously, as already discussed in Section \ref{sec:clouds} and illustrated in Figure \ref{fig:clocks}. Consequently, an inferred high-redshift $f_{\rm esc}$ contains two linked but distinct pieces of physics, namely the production of ionising photons by the stellar population and the transmission of those photons through the evolving gas distribution, further determining how escape fractions from different studies should be compared. An $f_{\rm esc,HII}$ of tens of percent and a galaxy-scale $f_{\rm esc}$ of a few percent are not necessarily contradictory measurements, because photons can escape the HII region and subsequently be absorbed in the diffuse ISM or CGM. Conversely, a substantial directly measured galaxy-scale LyC flux requires that sufficiently low-column-density channels persist beyond the individual HII regions and through the intervening ISM. The relevant quantities must therefore be compared at the same spatial scale and with the same definition of the intrinsic ionising photon budget. An escape fraction is consequently a measurement of how far ionising radiation has propagated through the multiphase gas before being absorbed.


\section{The engines: caveats of stellar population synthesis models}\label{sec:stellar_pops}
The majority of feedback-related quantities in Section~\ref{sec:obs_diag} require an assumption on the feedback-driving stellar population, unless the massive stars can be characterised with sufficient fidelity and completeness. The conversion from H$\alpha$ to $L_{\rm bol}$ and/or $Q(\mathrm{H}^{0})$, from a colour--magnitude diagram or an integrated spectrum to an age and SFH, and from an age to a supernova rate, all require choosing a model that describes (or prescribes) how a population of stars injects radiative and mechanical energy as a function of time. The resulting feedback quantity therefore inherits assumptions about stellar evolution, atmospheres, the IMF, binarity, rotation, mass loss, metallicity, and the star formation history. And because the stellar source term is the starting point from where feedback quantities are inferred, changing the predicted stellar population can change the inferred amount of radiation or mechanical energy available to interact with the gas, even when the observed gas properties are unchanged.

In very simple terms, a stellar population is synthesised from stellar evolution tracks (the path of a star of given initial mass and metallicity) by constructing isochrones (the stars present at a given age) and assigning each star a spectrum from an atmosphere library. The population can be of a single age, with all stars formed at the same time, or composite, with stars formed over a period of time. Under a chosen IMF, the synthesis then predicts the integrated light (colour, spectrum, luminosity) of the stellar population, which can be a star cluster or an entire galaxy. Full SED modelling requires combining the synthesised population with a star formation history, a metallicity history, nebular gas, and dust absorption and emission. Leading SPS codes include GALAXEV \citep{bruzual2003}, Starburst99 \citep{leitherer1999}, FSPS \citep{conroy2009,conroy2010a}, the Maraston models \citep{maraston2011}, and BPASS \citep{eldridge2017}.

SPS models are then used in two directions. In the forward (modeling) direction, an assumed IMF, set of evolutionary tracks and atmospheres, and SFH produce a synthetic spectrum or synthetic photometry for the adopted assumptions and a unique set of parameters --- $Q(\mathrm{H}^{0})$, $L_{\rm bol}$, wind power, and supernova rate --- as a function of age and metallicity. It therefore provides a controlled way of asking how a particular assumption about stellar physics changes the predicted observables or source terms. Inverse modeling starts from observed photometry or spectroscopy and attempts to recover the properties of the underlying stellar population; the inferred age, mass, star-formation history, attenuation, and metallicity are then used to estimate quantities such as $Q(\mathrm{H}^{0})$, $L_{\rm bol}$, $L_{\rm w}$, or supernova rate. The caveat is that the fit parameters are degenerate: more than one combination of age, star formation history, dust, metallicity, IMF sampling, rotation, and binarity can match the same photometry or spectrum, and the relative uncertainties of the input quantities propagate to the derived feedback quantities. If using SPS models, one should, at the very least, know what the chosen model assumes for stellar evolution and atmospheres, the IMF and the modelled mass range, the initial metallicity, binary parameters, and dust and nebular gas. We will now discuss the two directions of SPS model use separately.

\subsection{Forward modelling: predicting stellar, galaxy, and feedback properties from synthetic populations}\label{sec:sps_forward}
A population synthesis code takes an IMF, stellar tracks and atmospheres, and a star formation history, and returns the ionising photon rate, bolometric luminosity, and other quantities of interest in terms of feedback-related studies. Commonly used frameworks include Starburst99 \citep{leitherer1999}, FSPS \citep{conroy2009,conroy2010a}, and BPASS \citep{eldridge2017}. Different model families make different choices for the evolution of massive stars, mass loss, rotation and binary evolution, and consequently do not predict the same stellar source terms. For example, a comparison of BPASS, BoOST, Geneva, MIST and PARSEC models at near-solar metallicity found differences of up to 18\% in the integrated ionising radiation, while predictions for maximum stellar radii and remnant masses differed much more substantially \citep{agrawal2022}.

This makes forward modelling particularly useful for feedback studies because the sensitivity of a derived quantity to an uncertain stellar ingredient can be directly tested. The effect of changing the evolutionary tracks, including rotation or binary interactions, can be examined at fixed age, metallicity and IMF; stochastic sampling can be included explicitly for low-mass populations; and different wind prescriptions can be propagated into both the instantaneous wind power and the subsequent stellar evolution. In what follows, we briefly discuss how input choices affect derived  quantities of interest for feedback studies.

{\it Stellar wind mass loss.} The $\dot{M}$ inferred from diagnostics such as recombination lines depends on the assumed small-scale structure of the wind, and because the emissivity of recombination processes scales approximately as density squared, adopting a smooth instead of a clumpy wind can overestimate \(\dot{M}\) by factors of a few to ten \citep{puls2008}. Empirical clumping corrections therefore imply substantially lower O-star mass-loss rates than those obtained from smooth-wind prescriptions. These empirical constraints should, however, be distinguished from theoretical predictions of the line-driven wind mass-loss rate, which are parallel developments that also predict lower mass-loss rates than the canonical prescriptions used in many evolutionary calculations \citep{bjorklund2021,vink2022}. The magnitude of the systematic uncertainties depend on the adopted clumping prescription, wind diagnostic, stellar parameters, and metallicity. Importantly, a lower integrated mass-loss rate changes the evolutionary stripping of massive-star envelopes and can therefore alter the duration of Wolf–Rayet phases (and therefore the overall efficiency of WR-mediated feedback), stellar effective temperatures, remnant masses, and the subsequent ionising output of the stellar population \citep{vink2022,conroy2013}. Consequently, the choice of wind prescription affects not only the instantaneous wind power \(L_{\rm w}\), but also the predicted time-dependent \(Q({\rm \mathrm{H}^0})\) and, more generally, the stellar population spectrum. Wind uncertainties therefore propagate through both the mechanical feedback budget and the stellar property used to infer radiative feedback. 

The integrated mass loss also affects
whether a star ends its life as a successful CCSN or collapses to a black hole without a luminous explosion (e.g., \cite{sukhbold2016,oconnor2011}), by setting the final core mass and structure (e.g., the core compactness). Because explodability is not a monotonic function of initial mass, wind prescriptions affect not only the timing but also the number of CCSNe per unit stellar mass formed, and therefore the SN energy and momentum budget available for feedback. Furthermore, steady-wind prescriptions do not capture episodic mass loss which can eject substantial mass (e.g., $\gtrsim$ 10 M$_{\odot}$ ejected by $\eta$ Carinae's Great Eruption, \cite{smith2014}) and may account for a substantial fraction of the total mass lost by the most massive stars. 

{\it Binarity and rotation.} A large fraction of massive stars are expected to interact with a companion during their evolution. For example, \citet{sana2012} inferred a binary interaction fraction of more than 70\% for their sample of massive stars. Mass transfer and envelope stripping produce hot, luminous stars that continue to emit ionising photons after the ionising output from single stars has declined, becoming distinct around 6--8~Myr \citep{eldridge2017,scheuermann2026} and persisting for tens of Myr \citep{gotberg2019}. That delay falls after the typical CO--H$\alpha$ overlap and after the usually short obscured 21~$\mu$m phase of Section~\ref{sec:clouds}. Binary-stripped stars therefore add ionising photons once natal CO-bright gas has already been cleared, coinciding in time with the supernova tail of Figure~\ref{fig:clocks} and with the 5--25~Myr population that dominates DIG ionisation \citep{mcclymont2024}. Similarly, rotating stellar models can alter the ionising output relative to non-rotating models by up to a factor $\sim 5$ at solar metallicity \citep{leitherer2014}. The resulting change in the stellar source term can therefore affect interpretations of both nebular leakage and the timing of feedback relative to molecular-cloud dispersal.

{\it IMF sampling.} For individual low-mass associations the upper IMF is sampled stochastically rather than continuously. Deterministic SPS models then describe the mean population rather than a particular realisation. This is especially important for quantities dominated by the most massive stars, such as $Q(\mathrm{H}^{0})$. Stochastic population synthesis calculations show that the relation between $L_{\mathrm{H}\alpha}$ and $L_{\rm bol}$ can have substantial age- and mass-dependent scatter in populations of \(10^3\)--\(10^{4.5}\,M_\odot\) \citep{dasilva2012,rowland2024}. This is the same effect discussed in Section~\ref{sec:halpha}, i.e., applying a fully sampled IMF calibration to an individual region can produce a biased estimate of the stellar source term.

The important point is that these effects do not represent a single generic uncertainty introduced from the adopted SPS model, and their importance depends on the feedback quantity and on the timescale being probed. Rotation and IMF sampling can strongly affect the ionising luminosity of a young population, mass-loss prescriptions affect both wind power and subsequent stellar evolution, and binary evolution becomes particularly important once stripped stars contribute to the ionising spectrum at later times. Forward modelling therefore provides a useful experiment that can be used to identify which assumptions control a particular feedback inference.

\subsection{Inverse modelling: deriving stellar, galaxy, and feedback properties from synthetic populations}\label{sec:sps_inverse}
The inverse step fits an observed SED, ultraviolet continuum, or recombination-line luminosity with an SPS model to infer quantities such as stellar age, mass, star-formation history, metallicity and attenuation. The inferred population is then used to estimate the stellar source terms $Q(\mathrm{H}^{0})(t)$, $L_{\rm bol}(t)$, and $L_{\rm w}(t)$ from the time-dependent output of a population synthesis library, and from those one can then derive feedback-related quantities discussed earlier in this paper.

The main issue of deriving physical parameters by fitting observations to SPS models is that there is no unique combination of model parameters that matches the observation \citep{conroy2013}, so degeneracies enter the feedback measurements of Section~\ref{sec:obs_diag}. The age of a stellar population is the most immediate, and this is illustrated in the PHANGS HII region--association matches of \citet{scheuermann2026}, which have a median age of 2~Myr, and $Q(\mathrm{H}^{0})$ falls with time. They show that a 1~Myr uncertainty can halve the derived $Q(\mathrm{H}^{0})$, and that this is the largest term in the propagated error on $f_{\rm esc}=(Q(\mathrm{H}^{0})-Q_{{\rm H}\alpha})/Q(\mathrm{H}^{0})$. At a given H$\alpha$ luminosity, that factor of two in the stellar ionising rate is a factor of two in inferred leakiness, and in the input used to compute $P_{\rm dir}$. That is, an older fitted age lowers $Q(\mathrm{H}^{0})$ and $f_{\rm esc}$ and can make the derived stellar content appear insufficient to power the nebula, while a younger age inflates both leakiness and radiation pressure. Even at fixed age, the same HII region--association pairs yield different $Q(\mathrm{H}^{0})(t)$ under different libraries (e.g.~w/o rotation, binary evolution) \citep{scheuermann2026} and IMF sampling assumptions. Ultimately, the choice of which models to use should therefore be guided by the scientific question one is asking.

\vspace{0.5 cm}
Nearby galaxies are ideal laboratories to ground-truth SPS models because individual massive stars and clusters can be resolved. Photometry of those stars, or of the population colour--magnitude diagram, can be compared to evolutionary tracks to derive temperatures, luminosities, and, for an association or cluster, a mass and an age. Spectroscopy supplies spectral types, luminosity classes, and wind diagnostics, which atmosphere grids convert into $Q_{0}$, $L_{\rm bol}$, and $\dot{M}$ star by star. Summing over the known population of stars and their individual parameters then gives an empirical prediction that can be compared with an integrated-light SPS fit of the same region, and with the nebular recombination luminosity \citep{mcleod2019,mcleod2020,scheuermann2026}. Completeness, crowding, and the atmosphere grid used to assign $Q(\mathrm{H}^{0})$ to a spectral type do not fully remove model assumptions and uncertainties, but such comparisons test whether the population assumed in the feedback inference is consistent with the resolved stellar content. At high redshift, individual (massive) stars and star clusters generally cannot be resolved, so integrated spectra provide no equivalent star-by-star constraint, which is precisely why SPS models need to be calibrated with resolved stellar populations, therefore providing a direct empirical test of the stellar properties assumed by feedback quantities derived from integrated light.


\section{Stellar feedback at high redshift}\label{sec:highz}
The feedback diagnostics of Section~\ref{sec:obs_diag} and the stellar models of Section~\ref{sec:stellar_pops} were largely calibrated on nearby galaxies. Applying them at high redshift, where star-forming galaxies are characterised by lower metallicities, higher gas fractions, and higher star formation rates \citep{maiolino2019,tacconi2020}, therefore comes with two questions. First, whether stellar feedback couples to the ISM differently under the physical conditions of the early Universe, and second, even if the underlying feedback physics does not change at high redshift, do the conversions from observables to stellar source terms and ISM properties as discussed in Section \ref{sec:obs_diag} remain valid. 

{\it JWST} has made these questions urgent by delivering rest-frame optical photometry (and subsequent spectroscopy) of galaxies within the first billion years. \citet{labbe2023} identified 13 very red candidates at photometric $6.5<z<9.1$ whose spectral energy distributions show two sharp breaks, interpreted as a Balmer break plus a Lyman break. Six have fiducial stellar masses above $10^{10}\,M_\odot$, taken as the median across several SED codes, under the assumption of a Salpeter IMF. If the derived masses are taken at face value, the interpretation is that the stellar-mass density in galaxies above $10^{10}\,M_\odot$ exceeds prior UV-selected expectations by a factor of $\sim 20$ at $z\sim 8$ and $\sim 1000$ at $z\sim 9$. In addition to this, \citet{finkelstein2024} constructed rest-UV luminosity functions for 88 $8.5<z<14.5$ galaxies and found that the space density at $M_{\rm UV}=-20$ changes only modestly from $z\sim 14$ to $z\sim 9$. Subsequent NIRSpec spectroscopy has confirmed that the UV-luminous population at $z\sim 10$--14 is, with exceptions, at the claimed redshifts \citep{curtislake2023,arrabalharo2023,carniani2024} (although these objects have stellar masses of order $10^{8}$--$10^{9}\,M_\odot$, compared to the higher masses of the Labb\'e sample, for which a spectroscopic high-mass analogue exists albeit at a lower redshift of $z\sim 5.5$, \citep{xiao2024}). These results imply rapid early assembly of stellar mass and of UV-luminous sources, which could either be the result of weaker feedback regulation, or of a higher stellar yield (i.e., more ultraviolet light and ionising photons per unit stellar mass produced by a top-heavy IMF or very massive stars). Although the two options might seem to contradict each other, whether a higher stellar yield translates into stronger feedback is a question of coupling. Weaker regulation by stellar feedback would mean that the high masses are real, whereas a top-heavy IMF or a helium-boosted mass-to-light ratio would mean that a Salpeter-based SED overestimates the derived stellar mass.

Even where redshifts are spectroscopically confirmed, the derivation of stellar masses of these sources still depends on assumptions about how much light a stellar population produces per unit mass. For example, a top-heavy IMF, or a larger contribution from very massive stars, would increase the UV and ionising output per unit mass assembled. Helium enrichment at the levels inferred for some nitrogen-rich $z\gtrsim 6$ systems can boost the 1500~\AA\ luminosity of a zero-age main-sequence population by up to $\sim 50\%$ \citep{katz2024}. An altered upper mass function and a helium-enriched composition enter the mass-to-light ratio of Section~\ref{sec:stellar_pops}, and therefore the $M_\ast$ and the UV luminosity inferred from a given SED. If a standard IMF and composition are assumed, assembling more or less stellar mass at a given gas reservoir requires a change in how efficiently stellar radiation and winds couple to the ISM, which in turn is tied to the ISM properties, such as the ISM density of the star-forming regions. 

In Section \ref{sec:obs_diag} we have seen that compact, high-column HII regions can trap ionising radiation and reradiate it in the infrared until radiation pressure on dust exceeds the ionised-gas pressure, as in the Galactic UCHII sample of \citet{olivier2021}. These probe a different regime compared to the PHANGS HII region population, where \citet{pathak2025} find a median direct-radiation trapping factor $f_{\rm trap}\approx 0.05$ and the pressure of the ionised gas typically exceeds the direct radiation pressure. High-redshift star-forming galaxies are gas-rich compared to nearby systems \citep{tacconi2020}. Hence, under the assumption that their star-forming regions are also compact and high-column, high-z galaxies are likely closer to the trapped-infrared regime (although a high gas fraction alone does not imply stronger IR trapping, which requires a higher gas surface density, dust opacity, and also depends on geometrical/compactness properties). Therefore, what is required is the knowledge of the density, compactness, and infrared reprocessing in the high-redshift systems. 

It needs to be said that additional assumptions and uncertainties remain, such as the contribution of AGN or contamination of nebular emission, or the assumption about a burst vs continuous SFH. To illustrate this further, \citet{cameron2024} show that the SED of GS-NDG-9422 at $z=5.943$ fitted as starlight can be attributing nebular light to stars because nebular two-photon and free-bound emission can dominate the rest-UV and produce a Balmer jump. The apparent excess can therefore arise through at least three branches, namely, a different stellar yield per unit mass, different coupling of that output to the ISM, or an inadequacy of the diagnostic model used to infer the stellar population.

At high redshift, the conversions of Sections~\ref{sec:obs_diag} and~\ref{sec:stellar_pops} are applied to integrated light, without knowledge of the underlying stellar population, whereas in nearby galaxies we have the benefit of resolved photometry and spectroscopy of individual massive stars which, as mentioned in Section \ref{sec:stellar_pops} can be used to calibrate SPS models. Until that calibration, or ground-truthing, has been performed, age, IMF, binary evolution, and stellar atmospheres remain priors on the inference of feedback quantities derived from high-z observations. Whether radiation and winds couple more or less efficiently to the ISM as a function of redshift is not currently known. Local studies do show that the conditions of the early Universe (such as lower metallicities, higher gas fractions, and higher star formation rates) affect how stars form and how much radiative and mechanical feedback a population produces. Again, the practical care one needs to take is therefore the choice of the SPS model and the ISM assumptions (extinction, density, aperture) when a high-redshift spectrum is converted into a feedback quantity. A good example for this comes from analyses of deep JADES spectra at $z\sim 6$--9 that show how the high ionisation parameters inferred from the strong line ratio $\mathrm{O}_{32}$ depend on the adopted stellar library and photoionisation grid \citep{cameron2023}. 

The insight gathered from the latest high-redshift results is therefore that {\it JWST} has exposed how strongly the inferred stellar source term and ISM properties depend on the adopted stellar and nebular models more than it has shown that a particular feedback channel changes with redshift. To test whether and how feedback it couples to the ISM changes with redshift requires observations that constrain both the stellar output and the physical conditions into which that output is injected.

\section{The role of environment}\label{sec:env}
As has become clear through the discussion in the preceding sections, stellar feedback needs to be contextualised in terms of environment, that is, metallicity (and therefore dust content), ISM conditions (density, temperature, multi-phase gas reservoirs), but also more global properties such as galaxy (stellar) mass, star formation rate, and dynamics. These environmental properties impact both the feedback physics and the coupling efficiency to the ISM. We first treat metallicity and dust, which jointly modulate the stellar output and the gas that absorbs it. We then discuss larger galaxy-scale properties in terms of whether the energy and momentum are retained at the local scales they are injected in.

\subsection{Metallicity and dust content}
Metallicity impacts both stellar and gaseous physics \citep{geen2023}. Decreasing metal abundance weakens line-driven winds \citep{vink2022}, while typically raising the ionising photon emission rate per unit stellar mass, because the atmosphere absorbs fewer photons. In the gas, cooling by metal lines is less efficient at low $Z$, which can enhance the impact of mechanical and photoionisation feedback. Dust also affects how much of the stellar radiative momentum is transferred to the gas \cite{geen2023}, but this does not enter the observed $P_{\rm dir}$ directly, with $L_bol$ in Eq.~\ref{eq:pdir} inferred from H$\alpha$ already excluding escaping ionising photons, but assuming that non-ionising photons are absorbed in the same proportion. Ionising photons deposit their momentum inside the ionisation front even without dust, where it is balanced by a pressure gradient within the ionised gas \cite{draine2011}. Non-ionising photons, which carry most of bolometric luminosity, are absorbed by dust only, both within the ionised gas (to an extent set by its dust optical depth) and in the surrounding neutral shell. 

\citet{rowland2024} provide a continuation of the study of the relative importance of feedback-related pressure terms in the more massive and higher metallicity PHANGS galaxies and the conditions of the Milky Way and Magellanic Clouds into the dwarf ($M_\ast\sim 4\times 10^7\,M_\odot$), low-$Z$ ($12+\log({\rm O/H})\sim 7.5$) regime. They find that also in these systems the ionised gas pressure exceeds direct radiation pressure, with lower $P_{\rm dir}$ values at lower metallicities. This is expected when averaged across different systems (ignoring systemtatics stemming from different analysis methods) where gas-phase metallicity and dust content are expected to positively correlate when taken over a large dynamic range of the two \citep{remy2014}. Note that the lower values of $P_{\rm dir}$ measured at low metallicity do not imply that radiative feedback as a whole becomes less important: a lower dust-to-gas ratio reduces the destruction of trapped Ly$\alpha$ photons and raises the force multiplier mentioned in Section \ref{sec:definitions}. Theoretical work therefore predicts that the pressure of Ly$\alpha$ photons increases towards low metallicities (e.g., \citet{kimm2018}), while the observed $P_{\rm dir}$ decreases. Strictly speaking, the momentum budget as a whole in the low-metallicity regime is therefore observationally incomplete. Combining neutral-gas columns (e.g., from HI 21 cm or Ly$\alpha$/metal absorption) with resolved Ly$\alpha$ spectroscopy of nearby metal-poor HII regions would provide the first empirical constraints on this term.

But resolved radial trends in nearby galaxies show that dust and metallicity do not always exhibit the same behaviour. For example, the inner ($<0.5$ R$_{e}$) parts of the NGC~300 show a negative metallicity gradient but a positive dust gradient, along which \citet{mcleod2021} find that both $P_{\rm ion}$ and $P_{\rm dir}$ increase weakly. In absolute terms, the ionised gas pressure dominates over the direct radiation pressure in this portion of the galaxy, and the $P_{\rm ion}$ trend is the expected response to a harder spectrum and a higher photon flux at lower metallicities. The rising dust content in this region, seen both in the Balmer decrement and independently in far-infrared dust maps \cite{casasola2017}, coincides with the increase in $P_{\rm dir}$. Since $P_{\rm dir}$ does not depend explicitly on dust, this trend must act through $P_{\rm bol}$ or $r$, for example through more compact HII regions in denser, dustier gas.
We therefore caution against assuming that the dust follows the same behaviour as the metal abundance within individual systems. Across the inner few kiloparsecs of a disc, dust can remain constant or increase even as the gas-phase abundance falls, and that decoupling changes which pressure term is inferred to dominate.

\begin{figure}
    \centering
    \includegraphics[width=0.9\linewidth]{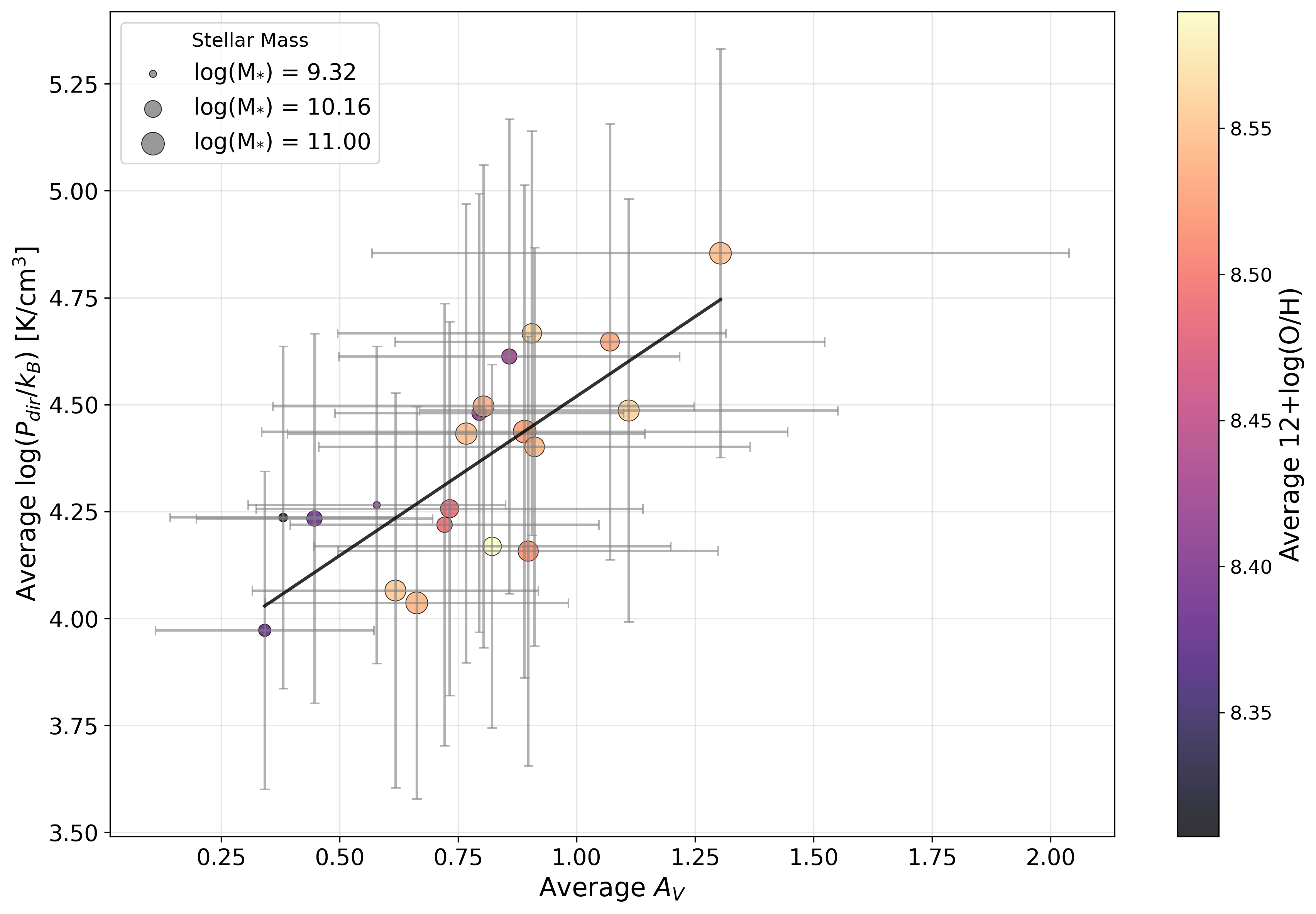}
    \caption{Correlation between the average direct radiation pressure $P_{\rm dir}$ and the average dust content (traced by the average dust extinction $A_{V}$) for publicly-available data of each of 19 PHANGS galaxies and NGC~300 (averaged are over all HII regions in each galaxy; black points, error bars correspond to the spread of values in each galaxy), and an unweighted straight line fit (black line). Data points are colour-coded by the average gas-phase metallicity, the marker size is a measure of the stellar mass of each galaxy.}
    \label{fig:average_av_pdir}
\end{figure}

Radial HII region pressure \citep{barnes2021} and metallicity \citep{brazzini2024} trends in PHANGS galaxies further show that these gradients can exhibit significant variation and structure within individual galaxies. In a forthcoming analysis, included here only as an illustration and therefore should not to be as an established literature result until published, McLeod et al. (in prep.) analyse how radial feedback and dust extinction trends are modulated by the large scale galaxy properties and morphology. These preliminary findings also show that, when averaged over entire galaxies rather than looked at in a resolved manner and therefore washing out any internal radial trend variations, nearby galaxies show a positive correlation between the direct radiation pressure and the dust content, again proxied by the average dust extinction, as shown in Fig.\ref{fig:average_av_pdir}, where dustier galaxies exhibit higher direct radiation pressures.

Dust also sets whether infrared radiation is trapped, which is an effect of compactness as much as metallicity. Compact Galactic HII regions can have $P_{\rm IR}/P_{\rm dir}$ of several to ten \citep{olivier2021,lopez2011}; PHANGS discs at $\sim 10$~pc have reprocessed-to-direct ratios of order 5\% \citep{pathak2025}. This difference is a result of the fact that the probed environments are different. Further, the mid-IR 21~$\mu$m emission already discussed in Section \ref{sec:clouds} is not only a tracer of embedded star formation, it is also a molecular-column tracer \citep{ramambason2026}, which is why the CO--21~$\mu$m clock of Section~\ref{sec:clouds} is not a pure star formation clock. One might then expect faster cloud clearing at low metallicity, because more ionising photons and less dust-shielded CO would shorten the overlap. \citet{ramambason2026} report a tentative trend toward faster 21~$\mu$m and obscured clearing at lower metallicity, but as also already mentioned, the sample on which these measurements were made only reaches $\sim 0.4\,Z_\odot$, therefore not probing the extremely low metallicity regime of dwarf galaxies. We caution that a causal connection between metallicity and feedback timescales and cloud clearing, while tempting, cannot be made uniquely from any of the above until the properties of the stellar sources and the feedback terms have been derived in a consistent manner for all HII regions in the probed galaxies.

So far, the discussion of varying metal and dust content environments has focused on radiative feedback. While available for HII regions in PHANGS galaxies \citep{barnes2021}, and few tens of Galactic UCHII \citep{olivier2021} and LMC and SMC regions \citep{lopez2014,mcleod2019}, mechanical wind feedback measurements are limited and a trend with metallicity has not been established yet. Other than limited measurements, a comparison between the different values is hampered by the fact that different measurement techniques (X-ray emission vs terminal wind velocity arguments) have been used. However, the weaker winds of lower metallicity stars would lead to an expected trend where $P_{\rm w}$ is negatively correlated with metallicity. 

Moving from pre- to SN feedback and its dependence on metallicity, an indirect empirical argument for enhanced SN feedback comes from \citet{mcleod2021}
who attempt to connect the above discussed HII region trends to the density of the medium into which supernovae explode. They infer pre-shock ISM densities of a few cm$^{-3}$ for seven supernova remnants and find a negative correlation with galactocentric radius and a positive correlation with gas-phase abundance, each at roughly $2\sigma$ once uncertainties are folded in. The interpretation is that supernovae at larger galactocentric distances, and therefore lower $Z$, expanded into less dense gas, as expected if stronger pre-supernova feedback had already carved cavities \citep{lucas2020}. The tentative conclusion would be that this enhances the effect of SN compared to a scenario in which SNe explosions are confined by higher density gas. However, the sample of supernova remnants in this study is small, and the radial coverage does not cover the galactic outskirts beyond $0.45R_{25}$. This type of measurement is therefore not predictive of a metallicity law for supernova coupling unless expanded to a much larger sample spanning a wider range of environmental properties.

Photon leakage from HII regions is equally impacted by the harder ionising spectrum and higher $Q(\mathrm{H}^{0})$ per unit stellar mass, and the smaller fraction of photons locally absorbed by less dust.
In the Dwarf Galaxy Survey, \citet{ramambason2022} infer HII region escape fractions that reach $\sim$60\% at low metallicity, are globally lower at higher metallicity, and correlate most strongly with specific star formation rate. Direct LyC detections in compact, metal-poor starbursts confirm that escape does occur in this regime \citep{izotov2016}, implying a higher cloud-scale escape fraction at low metallicities and low dust content, likely a consequence of local conditions. The mechanical implication is the one already suggested by the NGC~300 supernova remnants, that is, if radiative and mechanical feedback lowers the ISM density and open channels via which photons can escape before the first supernova, subsequent explosions couple to a more porous, lower-density medium \citep{mcleod2021,lucas2020,fielding2018}.


\subsection{Global environmental properties: stellar mass, star formation rate, and dynamics}
On larger scales, global galaxy properties such as stellar mass, star formation rate, molecular gas content, and galactic dynamics modulate the conditions into which various feedback mechanisms are launched. More massive, more actively star-forming discs are typically more metal-rich and dustier \citep{remy2014,maiolino2019}, and they set whether the local pressure terms, supernova coupling, and photon leakage discussed above become galaxy-scale outflows and, potentially, net gas removal. The relevant questions are therefore: how do galaxy-averaged feedback pressures scale with these global properties; how does the supernova rate, and the density of the medium into which supernovae explode, change; and under which of those conditions is gas actually launched from the star-forming layer.

\begin{figure}
    \centering
    \includegraphics[width=0.9\linewidth]{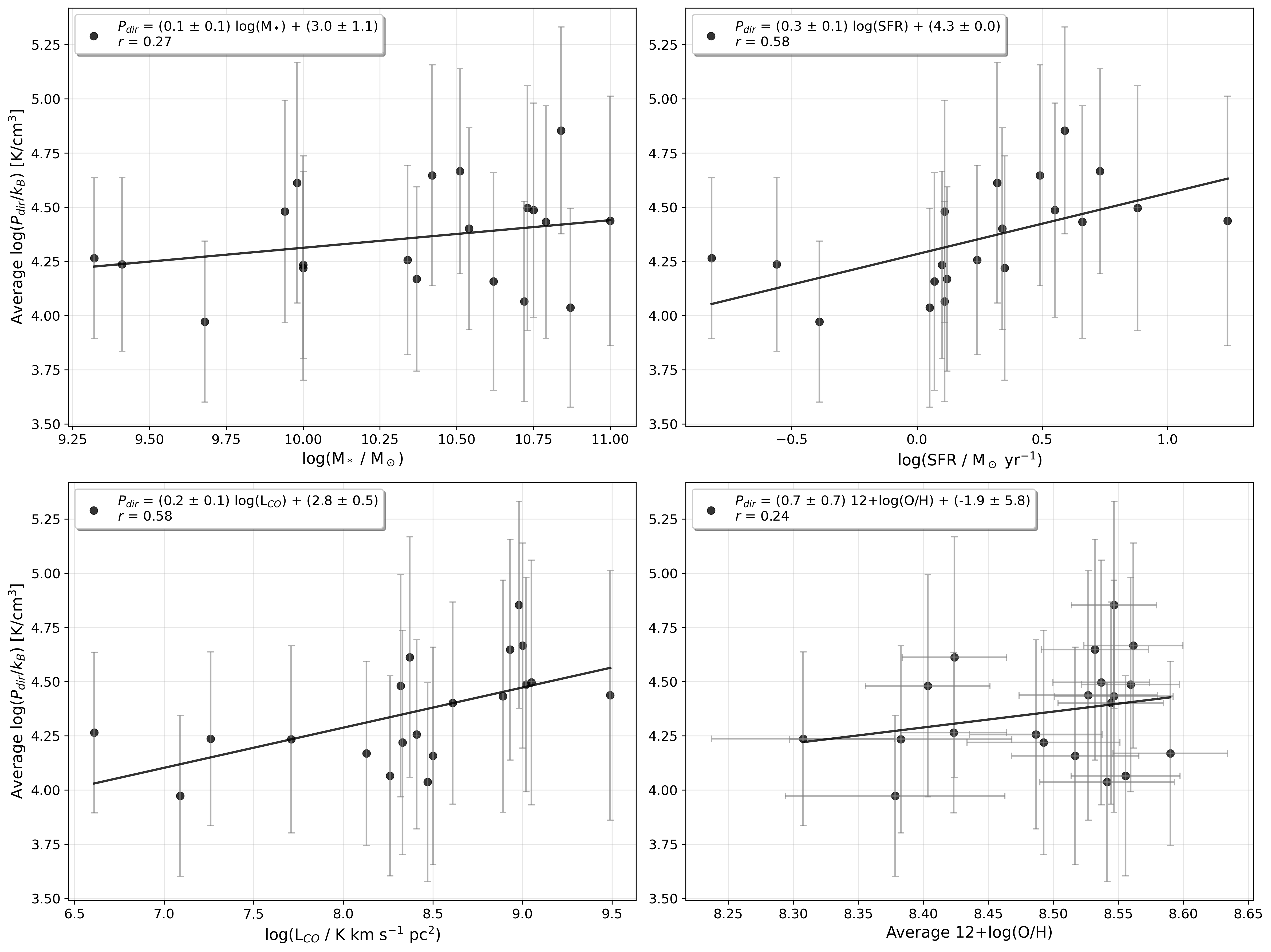}
    \caption{The direct radiation pressure $P_{\rm dir}$ averaged across all HII regions across 20 nearby galaxies (same as Fig.~\ref{fig:average_av_pdir}), as a function of global galaxy properties (from upper left to lower right: stellar mass, SFR, $L_{CO}$, and metallicity). Black lines correspond to linear fits with best-fit parameters listed in each panel's legend, together with the correlation coefficient.} 
    \label{fig:glob_prop}
\end{figure}

When averaged over entire galaxies, the same HII region measurements already discussed at length in Section \ref{sec:halpha} show that $\langle P_{\rm dir}\rangle$ also rises with galaxy stellar mass, star formation rate, and CO luminosity, as illustrated in Figure \ref{fig:glob_prop}. But because these global properties covary with each other, the trends seen in Figure \ref{fig:glob_prop} do not distinguish which property is their dominant driver. This is, however, a preliminary empirical link from the local pressure budget to the galaxy-scale quantities.

Having established that the ISM into which those supernovae explode is itself a function of pre-supernova radiative and mechanical feedback (and therefore of metallicity, dust, etc.), and having discussed the fact that the relatively short (of the order of a few Myr) CO--H$\alpha$ overlap favours pre-supernova clearing of CO-bright gas in nearby discs we now relate pre-SN and SN feedback to global environmental properties, knowing that galaxies with higher SFR produce more core-collapse events per unit time \citep{madau2014}. At lower metallicities and lower dust, and therefore in systems with lower stellar masses, one would expect radiative pre-SN clearing to be at least as efficient, if not more so, because of the ionising output per unit stellar mass can be higher and their dust shielding weaker. Whether the net pre-SN clearing efficiency increases toward lower mass therefore remains an empirical question, also because weaker winds from low-metallicity stars act in the opposite direction. But if pre-SN clearing is enhanced, a substantial fraction of supernova energy can vent into the halo rather than remaining stored in the bubble \citep{fielding2018}. In the opposite regime of high stellar mass, high SFR, high molecular-gas content, more supernovae explode, but they do so in dustier, more compact, and more often confined environments (centres, bars) where radiative losses and stalling are more likely before breakout \citep{barnes2020,dellabruna2022}. Galaxy dynamics enter the confinement picture with bars and shear, which modulate how long CO remains visible.
Outflow launching and net gas removal follow from this combination of dynamics, local pressure, supernova coupling, and potential-well depth. In shallow potentials a given energy injection more readily expels gas \citep{dekel1986}, an argument that contributes to the fact that dwarf galaxies are prime loci to study star formation-driven, galaxy-scale outflows. \citet{marasco2023} find, for a sample of local dwarf starburst galaxies, that the mass outflow rates and mass loading factors are low and strongly correlate with stellar mass, sSFR, and  \(\Sigma_{\rm SFR}\), concluding that in these galaxies feedback does not drive large-scale, mass-removing outflows.


The metallicity and dust trends therefore affect the two ends of the mass sequence differently. Low-mass, high-SFR, low-metallicity, low-dust systems combine a leakier ISM, weaker stellar winds, a harder ionising spectrum, and supernovae that may explode into pre-cleared cavities, in a shallow potential. These conditions favour both photon escape and mechanical venting, which is why these galaxies are often invoked as laboratories for LyC-escape studies \citep{izotov2016,ramambason2022,rowland2024}. High-mass, high-SFR, high-$L_{\rm CO}$ discs combine a larger supernova rate and (on average) higher direct radiation pressure with more dust trapping, more nuclear confinement, and a deeper well. Whether that configuration launches a hotter, more mass-loaded wind, a fountain, or a stalled nuclear flow depends on whether the extra pressure and the extra supernovae couple before the gas cools, mixes, or is reaccreted.

We suggest that progress in this area requires observations where $\eta$ is measured across systems at fixed phase (atomic, molecular, ionised) and radius, while controlling for stellar mass, SFR, dust, and metallicity. Outstanding questions that can already be addressed with the existing surveys referenced above are: (i)~once stellar mass, SFR, and $L_{\rm CO}$ are controlled for, does the galaxy-averaged $P_{\rm dir}$ still correlate with dust, or does one of the global quantities dominate; (ii)~does the confined HII region fraction remain $\sim$1\% outside galactic nuclei once bars and arms are separated, and do supernovae in those confined regions expand into systematically denser gas than in the disk; (iii)~can warm, molecular, and hot $\eta$ be measured in dwarfs and larger spirals and compared at matched $M_\ast/M_{\rm halo}$ after cloud and feedback timescales have been measured? Answers to these questions will shed light on how local coupling becomes galaxy-scale gas removal.

\section{Conclusions}\label{sec:conclusions}
The central questions of empirical stellar feedback studies are how much energy and momentum is intrinsically injected by the stars, how that energy and momentum are redistributed through the multiphase ISM, on what timescales this occurs, and which fraction remains dynamically effective on the different scale of interest. The observable consequences of feedback therefore depend as much on the properties of the surrounding feedback-driven gas as on those of the feedback-driving stars themselves. Several conclusions can be drawn from the discussion.

\begin{itemize}
    \item First, feedback quantities are model-dependent conversions, or inferences, from measurements. While feedback-related pressure terms, escape fractions, and mass-loading factors are physically useful quantities, their numerical values have meaning only together with the assumptions used to derive them.
    \item Second, the response of the ISM is a multi-phase, multi-scale, multi-temporal, and therefore multi-wavelength, phenomenon, and a diagnostic that samples one phase, one scale or one evolutionary stage cannot, by itself, constrain feedback.
    \item Third, the timescale of the response of the impacted ISM is as important as the magnitude of the feedback effect. And while the few-Myr spatial overlap between CO and tracers of young massive stars in nearby galaxies favours substantial evolution of the CO-bright phase driven by pre-SN feedback mechanisms before the first core-collapse supernovae, the tracer overlap does not distinguish which the physical mechanism is responsible, or dominating, the pre-SN clearing. Further, the subsequent supernova phase, binary-stripped stars, and ionisation of the diffuse gas extend the feedback sequence well beyond the typical lifetimes of molecular clouds.
    \item Fourth, the stellar source term is itself uncertain. Forward stellar population synthesis modeling shows how assumptions about stellar evolution, mass loss, rotation, binary evolution, IMF sampling, and metallicity affect the predicted ionising and mechanical output. Nearby observations of resolved stellar populations and the ISM provide an empirical route to testing and ground-truthing SPS models so that they can be used for robust inverse modeling, which otherwise adds uncertainties stemming from degeneracies in the model parameters. This calibration becomes increasingly important at high redshift, where the underlying stellar population cannot be resolved and where {\it JWST} observations probe regimes in which local stellar population and nebular calibrations may not apply unchanged.
    \item Finally, feedback and environment are interdependent. Metallicity, dust, compactness, gas density, and a galaxy's stellar mass, star formation rate and dynamics alter both the amount and spectrum of feedback produced by the stars and the medium into which it is injected. A change in $P_{\rm dir}$, $P_{\rm ion}$, $f_{\rm esc}$, or $\eta$ with galaxy mass or redshift can therefore arise from a change in the stellar source term, a change in the ISM structure, a change in the coupling, or some combination of all three. Separating these possibilities requires measurements of the source and response within the same systems and with consistent assumptions.
\end{itemize}

The observational picture that emerges from the preceding sections suggests an extension of the traditional {\it feedback ladder} \citep{bally2011}. The ladder provides a useful ordering of feedback mechanisms by their characteristic strength and the stellar populations and spatial scales on which they become important. Increasingly resolved, multi-phase observations, show that the effect of a given feedback mechanism cannot be interpreted independently of the environment in which it operates, because the same stellar energy or momentum input can produce different responses depending on the compactness, surface density, metallicity, geometry, gravitational potential, and larger-scale context of the surrounding gas. Its observational signature also depends on the spatial scale and temporal window over which the star/ISM system is measured. We therefore propose {\it contextualised feedback} as a framework in which feedback measurements are interpreted jointly in terms of the mechanism, environment, spatial scale, temporal scale, and gas phase being traced. In this view, a feedback measurement is a measurement of a feedback mechanism and its interaction with a particular physical system.

To end on a positive note, it is fair to say that the overall picture is relatively well constrained, at the very least more constrained than the apparent complexity of the problem might suggest: ionised gas pressure dominates many ordinary disk HII regions; radiation pressure becomes more competitive in compact, dusty environments; CO-bright gas generally ceases to overlap with young massive stars on few-Myr timescales; outflows are multiphase; and SPS assumptions affect inferred quantities. The most useful next step is therefore not simply to measure more feedback-related quantities in more regions, but to construct observations that cover the stars and the multi-phase ISM at the same spatial scales so that these can be independently characterised and the assumptions of the inference chains either minimised, or, at the very least, made sufficiently explicit. Moreover, while we have not discussed simulations, the real predictive power will result from observations and post-processed simulations that are matched around the same physical quantity, phase, and spatial scale, and analysed using the same methods. Only then can differences between observed and simulated feedback be interpreted as evidence for differences in physical coupling rather than differences in the inference chain.

\vspace{6pt}

\authorcontributions{Not applicable.}

\funding{This research was funded by the Science and Technology Facilities Council (STFC) consolidated grant ST/X001075/1.}

\institutionalreview{Not applicable.}

\informedconsent{Not applicable.}

\dataavailability{The data used to create figures \ref{fig:pressures}, \ref{fig:average_av_pdir}, and \ref{fig:glob_prop} is publicly available from the source publications \citep{rowland2024,lopez2014,mcleod2021,barnes2021,olivier2021,barnes2020}.}

\acknowledgments{During the preparation of this manuscript, the author used ChatGPT (OpenAI, GPT-5.6) for the purposes of language polishing to improve clarity and readability, and Gemini (Google) for the creation of Figure~\ref{fig:ladder}. The author retains full responsibility for the scientific content, interpretation, and final text of the manuscript.}

\conflictsofinterest{The author declares no conflicts of interest.}



\abbreviations{Abbreviations}{
The following abbreviations are used in this manuscript:
\\

\noindent
\begin{tabular}{@{}ll}
AGN & active galactic nucleus \\
ALMA & Atacama Large Millimeter Array \\
BC03 & \citet{bruzual2003} stellar population models \\
BPASS & Binary Population and Spectral Synthesis \\
CALIFA & Calar Alto Legacy Integral Field Area survey \\
CCSN & core-collapse supernova \\
CGM & circumgalactic medium \\
CMZ & Central Molecular Zone \\
COS & Cosmic Origins Spectrograph \\
CR & cosmic ray \\
DIG & diffuse ionised gas \\
DUVET & Deep near-UV observations of Entrained gas in Turbulent galaxies \\
EAGLE & Evolution and Assembly of GaLaxies and their Environments \\
ELT & Extremely Large Telescope \\
FIR & far-infrared \\
FIRE & Feedback In Realistic Environments \\
FSPS & Flexible Stellar Population Synthesis \\
FUV & far-ultraviolet \\
GMC & giant molecular cloud \\
HIM & hot ionised medium \\
HST & Hubble Space Telescope \\
IFU & integral field unit \\
IFS & integral field spectroscopy \\
IGM & intergalactic medium \\
IMF & initial mass function \\
IR & infrared \\
ISM & interstellar medium \\
JADES & JWST Advanced Deep Extragalactic Survey \\
JWST & James Webb Space Telescope \\
KCWI & Keck Cosmic Web Imager \\
LEGUS & Legacy ExtraGalactic UV Survey \\
LyC & Lyman continuum \\
MIRI & Mid-Infrared Instrument \\
MUSE & Multi Unit Spectroscopic Explorer \\
NIR & near-infrared \\
NIRSpec & Near-Infrared Spectrograph \\
PHANGS & Physics at High Angular resolution in Nearby GalaxieS \\
SAMI & Sydney-AAO Multi-object Integral field spectrograph \\
SED & spectral energy distribution \\
SFE & star formation efficiency \\
SFH & star formation history \\
SIGNALS & Star formation, Ionized Gas, and Nebular Abundances Legacy Survey \\
SIMBA & SIMBA cosmological galaxy formation simulations \\
SLUG & Stochastically Lighting Up Galaxies \\
SN & supernova \\
SNR & supernova remnant \\
SPS & stellar population synthesis \\
TIGRESS & Three-phase Interstellar Medium in Galaxies Resolving Evolution with Star Formation and Supernova Feedback \\
UCHII & ultra-compact HII region \\
UV & ultraviolet \\
VLT & Very Large Telescope \\
WIM & warm ionised medium \\
WNM & warm neutral medium \\
WR & Wolf--Rayet \\
YSO & young stellar object \\
\end{tabular}
}


\begin{adjustwidth}{-\extralength}{0cm}

\reftitle{References}


\bibliographystyle{plainnat}
\bibliography{references}



%


\PublishersNote{}
\end{adjustwidth}
\end{document}